\documentclass[12pt,preprint]{aastex}
\usepackage{graphicx,epsfig}

\shorttitle{On the Spectral Lags and Peak-Counts...}
\shortauthors{\v{R}\'{\i}pa et al.}

\begin{document}

\title{On the Spectral Lags and Peak-Counts of the
Gamma-Ray\\ Bursts Detected by the RHESSI Satellite}

\author{J. \v{R}\'{\i}pa}
\affil{
Charles University, Faculty of Mathematics and Physics, Astronomical Institute,
V Hole\v{s}ovi\v{c}k\'ach 2, CZ 180 00 Prague 8, Czech Republic\\
Institute for the Early Universe, Ewha Womans University, 11-1 Daehyun-dong,
Seoul 120-750, Korea}
\email{ripa@sirrah.troja.mff.cuni.cz}
\author{A. M\'esz\'aros}
\affil{
Charles University, Faculty of Mathematics and Physics, Astronomical Institute,
V Hole\v{s}ovi\v{c}k\'ach 2, CZ 180 00 Prague 8, Czech Republic}
\email{meszaros@cesnet.cz}
\author{P. Veres}
\affil{Department of Astronomy and Astrophysics, Pennsylvania State
University, 525 Davey Lab, University Park, PA 16802, USA\\
Konkoly Observatory, H-1505 Budapest, POB 67, Hungary\\
Department of Physics of Complex Systems, E\"otv\"os University,
H-1117 Budapest, P\'azm\'any P. s. 1/A, Hungary \\
Department of Physics, Bolyai Military University, H-1581
Budapest, POB 15, Hungary}
\email{veresp@psu.edu}
\author{I.H. Park}
\affil{Department of Physics and Research Center of MEMS Space Telescope,
Ewha Womans University, 11-1 Daehyun-dong, Seoul 120-750, Korea}
\email{ipark@ewha.ac.kr}

\begin{abstract}
A sample of 427 gamma-ray bursts from a database (February
2002 - April 2008) of the RHESSI satellite is analyzed statistically.
The spectral lags and peak-count rates,
which have been calculated for the first time in this paper,
are studied completing an earlier analysis of durations and hardness ratios.
The analysis of the RHESSI database
has already inferred the existence of a third group with intermediate duration,
apart from the so-called short and long groups.
First aim of this article is to discuss the properties of these intermediate-duration
bursts in terms of peak-count rates and spectral lags. Second aim is to discuss the
number of GRB groups using another statistical method and by employing the peak-count rates
and spectral lags as well.
The standard parametric (model-based clustering) and non-parametric
(K-means clustering) statistical tests together with the
Kolmogorov-Smirnov and Anderson-Darling tests are used. Two new results are obtained:
A. The intermediate-duration group has similar properties to the group of short bursts.
Intermediate and long groups appear to be different.
B. The intermediate-duration GRBs in the RHESSI and Swift databases
seem to be represented by different phenomena.
\end{abstract}

\keywords{gamma-ray burst: general}

\section{Introduction}

\citet{ma81,no84,kou93}; and \citet{apt98} have suggested the division
of gamma-ray bursts (GRBs) into two categories, either short or long,
according to their duration (at $\sim$\,2\,s). Many observations
demonstrate different properties of short and long bursts.
They have different redshift distributions \citep{ba06,os08}
and may have different celestial distributions \citep{bal98,bal99,mes00,lit01,mes03,vav08}.
At present there is a predominant opinion that they are physically different
phenomena \citep{nor01,bal03,fox05,kan11}.

There are also statistical indications of a third, ``intermediate", group.
The division of GRBs into three groups has been studied statistically over
different databases: BATSE \citep{ho98,mu98,bala01,ho02,ho06,cha07}; BeppoSAX \citep{ho09};
Swift \citep{ho08,hu09,ho10,ve10} and RHESSI \citep{rip09}.
These three groups may also have different celestial distributions \citep{mes00,vav08};
at least for the BATSE database.
No test has given a statistically significant support for
the existence of four or more groups.
Only the BATSE database gave a weak 6.2\,\% significance level
for such a possibility \citep{ho06}.

One cannot exclude an eventuality that the
separation of this third group
is simply a selection effect \citep{hak00,ra02}.
In other words, a separation from the statistical
point of view does not necessarily also indicate astrophysically different phenomena. In
principle, it is still possible that the class of the intermediate GRBs constitutes
a ``tail" of either the short or the long group. The article by \citet{ve10} claims that
- at least for the Swift database \citep{sak08} - the third group is related to the so-called X-Ray
Flashes (XRFs), which need not be physically distinct phenomena \citep{kip03,so06}.
Two models of XRFs are favored; either they are ordinary long GRBs viewed slightly off-axis
\citep{zhg02} or they are intrinsically soft long-duration GRBs \citep{gen07}.
Hence, at least in the Swift database, the problem of the
intermediate class seems to have been solved.

However, for three reasons the situation has not yet been clarified.
First, with regard to the Swift database,
another study suggests that even the short group
should be further separated \citep{sak09}.
Secondly, there is additional observational evidence
against the simple scheme that maintains the existence of
{\it only} two types of bursts (short/hard and
long/soft) separated at duration of $\sim$2\,s.
The GRB060614 event, which is clearly long at duration
($\simeq$\,100\,s) but in any other properties resembles a short GRB,
and subsequent short bursts with soft extended emission yet challenged this
scheme \citep{ge06}. To avoid the limitations of a short-long separation
terminology, the designations
``Type I" and ``Type II" have been proposed \citep{zhg06,zhg09,kan11} because
duration alone is hardly sufficient for a correct division into categories.
Thirdly, it remains possible that in other databases the
discovered intermediate group is not represented by XRFs.
Concerning this third reason,
the mean duration of the intermediate group appears to vary according to
the database in which it is found.
For the Swift data \citep{ho08,hu09} the mean duration is
$\sim$\,12\,s, which resembles the durations of the long GRBs, but for the
RHESSI and BATSE data \citep{ho98,mu98,ho06,rip09} this mean is far below 10\,s.

It is clear that any new result in the classification scheme of
GRB groups is desirable. In this article we
study the RHESSI database, where in addition to \citet{rip09},
the spectral lags and peak-counts are also included.
We have two concrete aims here:
First, to provide further statistical tests concerning the GRB classes and, second,
to provide additional information concerning the physical significance
of the RHESSI intermediate group found by \citet{rip09}.

The paper is organized as follows:
In Sec.~\ref{sec:sample} the RHESSI satellite and its GRB data sample are described.
In Sec.~\ref{sec:KS} distributions of spectral lags, normalized lags, and peak-count rates
are studied using Kolmogorov-Smirnov and Anderson-Darling tests along with Monte Carlo simulations.
In Sec.~\ref{sec:dis} we discuss results of these tests, compare the results with the BATSE and
Swift data samples, and discuss the number of GRB groups using model-based and K-means clustering methods.
Sec.~\ref{sec:sum} summarizes results of this paper.

\section{The RHESSI data sample}
\label{sec:sample}

The Ramaty High Energy Solar Spectroscopic Imager
\footnote{\url http://hesperia.gsfc.nasa.gov/hessi}$^,$\footnote{\url http://grb.web.psi.ch}
(RHESSI) is a satellite designed for the observation of hard X-rays and gamma-rays from solar
flares \citep{lin02}, but it is also able to detect GRBs.
There is no automatic search routine for GRBs,
and only if a message from any other instruments of the International Planetary Network (IPN)
occurs, the RHESSI data are searched for a GRB signal. Therefore, our data set includes
only events confirmed by other satellites.

In this paper we study the same list of bursts which has been published in
\citet{rip09}. We consider 427 GRBs from period between
February 14, 2002 and April 25, 2008.
Contrary to \citet{rip09}, the spectral lags and the peak-counts
- calculated for the first time for RHESSI -
next to the durations and hardnesses, are used.
They are collected, together with their uncertainties, in Table~\ref{tab:database}.
These new observational data allow further study of the questions concerning
the GRB classification. There are two arguments for the choice of the same list of bursts.
First, both in \citet{rip09} and in the present work, similar
statistical studies are performed. Hence, for comparison, it is reasonable to study
the structure of groups found in the RHESSI database over the same set.
The second argument concerns an instrumental effect.
The measurements of the hardness ratio of the events during the year 2008 and later has
been systematically affected by an ``annealing"
procedure\footnote{\url http://hesperia.gsfc.nasa.gov/hessi/news/jan\_16\_08.htm}
executed on the RHESSI detectors at late 2007 \citep{bel08}.
The reason why the RHESSI team decided to anneal the detectors was to recover its deteriorating
spectral sensitivity. However, the sensitivity at low energies had not been recovered as well as
at high energies and hence the measured GRB hardness ratios from the post-annealing period are
systematically shifted to higher values \citep{ver09,rip10}. In order to be eliminate this instrumental
influence a more sophisticated modeling would be required.
However, this is beyond the scope of this article.

In order to compare the spectral lags and the peak-counts of bursts belonging to the
different groups, one must provide a rule by which the particular GRBs are sorted into the concrete
groups. We proceeded in the following manner: The probability density function
employed in the fitting of the duration-hardness plane in \citet{rip09} is composed through the summation
of three bivariate log-normal
functions $f(x,y)=f_1(x,y)+f_2(x,y)+f_3(x,y)$, where $x$ is the base 10 logarithm of the duration
and $y$ is the base 10 logarithm of the hardness ratio, and $f_1$, $f_2$, and $f_3$ are components
corresponding to the particular groups. A burst at the point $[x_0;y_0]$ is considered short, intermediate
or long depending on whether the $f_1(x_0,y_0)$, $f_2(x_0,y_0)$ or $f_3(x_0,y_0)$ is maximal. In essence,
we follow a procedure identical to the that of
\citet{ho06} and \citet{ho10} utilized on the BATSE and Swift datasets.

In order to sort the given GRBs into the groups we employ
the measurements of the durations and hardness ratios as given
in Table~7 of \citet{rip09} with the exception of six events.
We found that for these six events the mentioned values in \citet{rip09} were
not corrected for a so-called decimation, which is an instrumental mode used to
conserve the onboard memory.
Table~\ref{tab:dec-corr} presents these six events, now corrected for this decimation.

\begin{figure}[h]
\centering
\includegraphics[trim=6mm 2mm 5mm 6mm,clip=true,width=0.6\columnwidth]{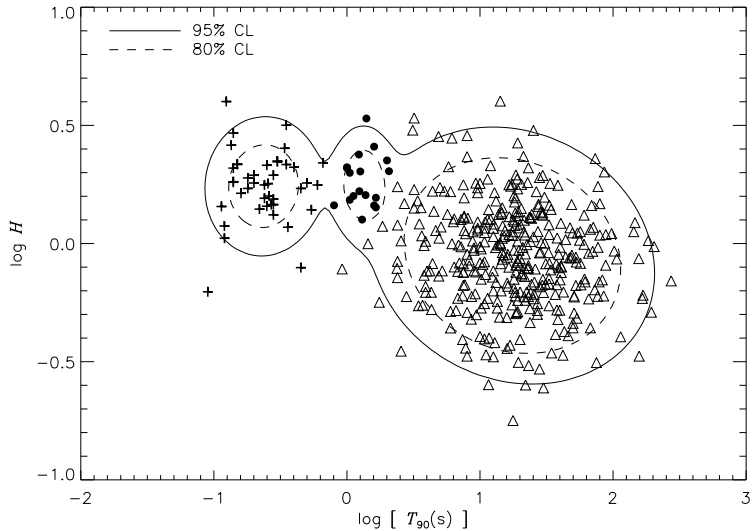}
\caption{The hardness ratio $H$ plotted against the duration $T_{90}$
for the RHESSI database with the best ML fit of three bivariate log-normal functions.
The different GRB group members are denoted with different symbols:
the crosses, full circles, and triangles correspond, respectively, to the short, intermediate,
and long bursts. CL means ``confidence level".}
\label{fig:groups}
\end{figure}

The group members used in this study were determined from the best Maximum
Likelihood (ML) fit \citep{rip09} in the duration-hardness plane of 427 GRBs. In this sample the
six events with corrected decimation were included along with the remaining
421 events taken from \citet{rip09}.
The best ML fit with
two bivariate lognormal components gives logarithmic likelihood $\mathrm{ln} L_2 = -313.4$.
Best fit with three components gives logarithmic likelihood $\mathrm{ln} L_3 = -303.4$.
The ML ratio test tells that the twice of difference in the logarithms of the likelihoods
$2(\mathrm{ln} L_3 - \mathrm{ln} L_2) = 20.0$ should follow $\chi^2$ distribution with 6
degrees of freedom \citep{ho06}. Therefore ML ratio test, employed in \citet{rip09} and now applied on
the duration-hardness plane with these 427 GRBs including the six events corrected for decimation,
gives again a statistically significant intermediate group at the significance level of 0.3\,\%.
The new (former) best-fit model parameters of the intermediate group are: 0.12 (0.11) for the mean
logarithmic duration, 0.25 (0.27) for the mean logarithmic hardness, 4.1 (5.3\,\%) for the weight,
and 0.0 (0.59) for the correlation coefficient. The group members are shown in Fig.~\ref{fig:groups}
and listed in Table~\ref{tab:database}.

\begin{deluxetable}{lccc}
\tabletypesize{\scriptsize}
\tablecaption{Six RHESSI GRBs with corrected $T_{90}$ durations and
hardness ratios.\label{tab:dec-corr}}
\tablewidth{0pt}
\tablehead{
\colhead{GRB\tablenotemark{a}} &
\colhead{Peak time\tablenotemark{b}} &
\colhead{$T_{90}$ (s)\tablenotemark{c}} &
\colhead{Hardness ratio log $H$\tablenotemark{d}}
}
\startdata
030518B & 03:12:23.050 & (1.86$\pm$0.07)E+1 & (2.90$\pm$0.27)E-1 \\
030519A & 09:32:22.500 & (3.20$\pm$0.27)E+0 & (5.31$\pm$0.61)E-1 \\
031024  & 09:24:14.350 & (4.30$\pm$0.17)E+0 & -(2.06$\pm$0.31)E-1 \\
040220  & 00:55:15.800 & (1.80$\pm$0.07)E+1 & (9.39$\pm$2.72)E-2 \\
050216  & 07:26:34.275 & (4.50$\pm$0.56)E-1 & (2.33$\pm$0.48)E-1 \\
050530  & 04:44:44.900 & (2.40$\pm$0.26)E+0 & (2.41$\pm$0.63)E-1 \\
\enddata
\tablenotetext{a}{RHESSI GRB number.}
\tablenotetext{b}{Peak time of the count light-curve in UTC.}
\tablenotetext{c}{The uncertainties were calculated
though the same procedure used in \citet{rip09}.}
\tablenotetext{d}{The hardness ratio was defined as the ratio of GRB counts
at two different bands, $H=S_{(120-1500)\mathrm{keV}}/S_{(25-120)\mathrm{keV}}$.}
\end{deluxetable}

\begin{figure}[h]
\centering
\includegraphics[trim=6mm 3mm 4mm 6mm,clip=true,width=0.8\textwidth]{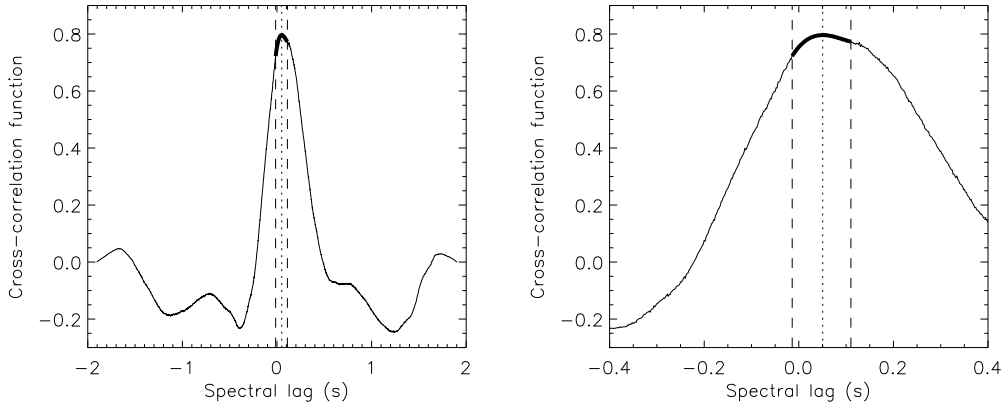}
\caption{\emph{Left}: An example of the cross-correlation function of two background-subtracted count
light-curves of the very bright GRB~060306 derived at two energy bands $400-1500$\,keV and $25-120$\,keV.
\emph{Right}: A detail of  the same curve with the third order polynomial fit (thick solid curve).
The position of
the maximum of the fit measures the spectral lag (dotted line).
The boundaries of the polynomial fit are marked with dashed lines.}
\label{fig:ccf}
\end{figure}

The spectral lags $L$ of the RHESSI data were calculated by ourselves
by fitting the peak of the cross-correlation function (CCF) of the
background-subtracted count light-curves at two channels, $400-1500$\,keV and $25-120$\,keV, by
a third order polynomial. The position of the maximum of the polynomial fit measures
the spectral lag. An example of such a fit is shown in Fig.~\ref{fig:ccf}.
The method is similar to that employed in the previous studies
\citet{nor00,nor02,fol08}; and \citet{fol09}
on the BATSE and INTEGRAL data. This is the first time that the spectral
lags have been calculated for the RHESSI GRBs.

To obtain statistical errors, a Monte Carlo (MC) method was utilized.
The following procedure was employed to prepare 1001 synthetic count profiles
for each GRB:
The measured count profiles were randomly influenced by Poisson noise,
after which the background was subtracted.
The RHESSI count rates are sometimes ``decimated'', which means that, as the rate becomes
too high or the onboard solid-state recorder becomes too full,
a part of the recorded counts is removed.
If decimation occurs, the fraction $(f_\mathrm{d}-1)/f_\mathrm{d}$
of the counts below a decimation energy $E_0$ is removed. $f_\mathrm{d}$ is the
decimation factor (weight), usually equal to 4 or 6. All events above $E_0$ are
downlinked\footnote{\url http://sprg.ssl.berkeley.edu/\textasciitilde dsmith/hessi/decimationrecord.html}.
To prepare the synthetic count profiles, the number of counts in each bin was changed accordingly
to the Poisson distribution. The 1-sigma errors for
non-decimated, fully decimated, and partially decimated data are
$\sqrt{C}$, $\sqrt{f_\mathrm{d}.C_\mathrm{dc}}$,
and $\sqrt{C_1+f_\mathrm{d}.C_\mathrm{2,dc}}$, respectively.
$C$ is the measured count number in a bin for non-decimated data.
$C_\mathrm{dc}$ is the count number in a bin of fully decimated data and consequently
corrected for this decimation.
$C_1$ is the count number in the non-decimated portion and $C_\mathrm{2,dc}$ is the corrected count
number in the decimated portion of the measured rate in the case of partially decimated data.
A detailed explanation is provided in Appendix~\ref{app:1}.
The CCF was fitted for each of the 1001 synthetic profiles and for each burst in our sample.
The median of such a distribution of 1001 maxima of polynomial fits
was taken as the true lag $L$ for each burst.
Theses median lags $L$ are used in the following statistical tests and listed in Table~\ref{tab:database}.
The 2.5\,\% and 97.5\,\% quantiles of such a distribution of 1001 maxima of polynomial fits for each GRB
delimit the 95\,\% CL statistical errors. These errors are also listed in Table~\ref{tab:database}.

We decided to calculate the spectral lags only for bursts with a signal-to-noise ratio
higher than 3.5 in both channels. This signal-to-noise ratio is
defined as $S_{\mathrm T90}/\sqrt{S_{\mathrm T90}+2B_{\mathrm T90}}$, where $S_{\mathrm T90}$ is a
GRB signal over the background level $B_{\mathrm T90}$, and both $S$ and $B$
are counts in a $T_{90}$ time interval over the range 25\,keV$-$1.5\,MeV. The choice of this limit was
found to ensure that the CCF was sufficiently smooth with a clear peak
allowing determination of a reliable lag.
Therefore, excluding the noisiest data, the number of GRBs with calculated lags is 142.
Their distribution is presented in Fig.~\ref{fig:lags}.

The GRB peak-count number $S$ was derived from the light-curve with the maximal count number $C$ at
the range 25\,keV$-$1.5\,MeV after subtracting the background $B$. The peak-count rate $F$ is given as
the peak-count number $S$ divided by the width of the time bin $\delta t_{\mathrm res}$. This width was
different for different GRBs, and covered a range between 2\,ms and 3\,s.
The dimension of the peak-count rate is count/s.

The one sigma error $\sigma_{\mathrm F}$ of the peak-count rate  $F$ was calculated as
$\sigma_{\mathrm F}=\sigma_{\mathrm S}/\delta t_{\mathrm res}$, where the error $\sigma_{\mathrm S}$ of
the GRB peak-count number is $\sigma_{\mathrm S}=\sqrt{(\sigma_{\mathrm C})^2+(\sigma_{\mathrm B})^2}$.
We assume that errors of the maximal count numbers $\sigma_{\mathrm C}$ and of the background
$\sigma_{\mathrm B}=\sqrt{B}$ are Poissonian and independent. The error $\sigma_{\mathrm C}$ is:
$\sigma_{\mathrm C}=\sqrt C$ in case of non-decimated data; given by expression (\ref{eq:full-dec})
in case of fully decimated data; and given by expression (\ref{eq:part-dec}) in case of partially
decimated data (see Appendix~\ref{app:1}). The peak-counts with errors were calculated for all 427 objects.

\section{Properties of the GRB groups}
\label{sec:KS}

\subsection{Distribution of spectral lags}
\begin{figure}[h]
\centering
\includegraphics[width=1.0\textwidth]{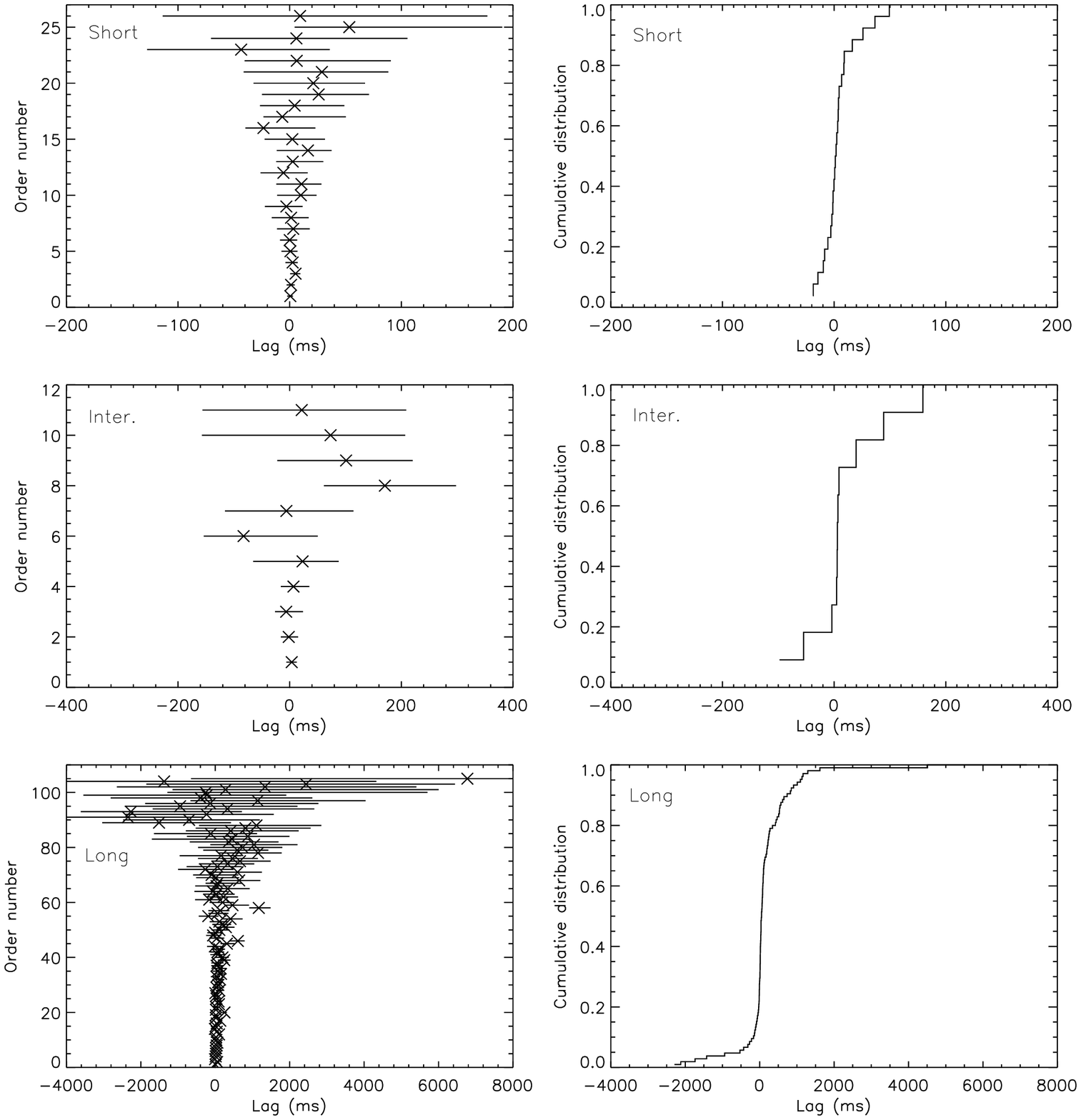}
\caption{\emph{Left panels:}
The spectral lags of RHESSI GRBs sorted along the $y$-axis with respect
to the value of (+error + $|$-error$|$) for short-, intermediate- and
long-duration bursts. The median lags, for each GRB, were taken from the
lags of 1001 synthetic background-subtracted count time profiles obtained
by Monte Carlo simulations of the measured profiles that were randomly
influenced by the Poissonian noise. The error bars are composed of the
95\,\%~CL statistical error and the profile time resolution. A positive
lag means that the low-energy counts are delayed. \emph{Right panels:}
The cumulative distributions of the obtained median lags for the three
groups of bursts are shown.
}
\label{fig:lags}
\end{figure}

In this section we use Anderson-Darling (A-D) test \citep{and52,dar57}
to compare distributions of spectral lags of different GRB groups (see Fig.~\ref{fig:lags})
found by the ML method applied on durations and hardness ratios (see Sec.~\ref{sec:sample}).
The short (intermediate, long) group contains 26 (11, 105) objects. The mean values of
the spectral lags of these groups are similar, hence we use the A-D test because it is
particularly sensitive to the tails of the tested distributions \citep{sch87}.
For its calculation we employ \emph{adk} package of the R software\footnote{http://cran.r-project.org} \citep{R}.
The results are summarized in Tab.~\ref{tab:lags}.

\begin{table}[h]
\centering
$
\begin{array}{cccc}
\begin{tabular}{ccc}
\hline\hline
Groups & A-D $P$ \\
       & (\%)    \\
\hline
Inter.-Short & 16.8 \\
Inter.-Long  & 4.2 \\
Short-Long   & $<10^{-3}$ \\
\hline
\end{tabular}
\begin{tabular*}{1.0cm}{c}
\\
\end{tabular*}
\begin{tabular}{cccc}
\hline\hline
Group &  Mean $L$ & Median $L$ & $\sigma$ \\
      &  (ms)     &  (ms)      &   (ms)   \\
\hline
Short  &   4.9  &  1.9  & 16.7  \\
Inter. &  28.7  &  5.9  & 78.4  \\
Long   & 178.0  &  50.8 & 874.9 \\
\hline
\end{tabular}
\end{array}$
\caption{
\emph{Left part:}
Results from the A-D tests of the spectral lag distributions for
the RHESSI database, are presented. The null hypothesis is that the two samples
are drawn from the same distribution. $P$ denotes the P-value of the test.
\emph{Right part:}
The means, medians and standard deviations $\sigma$ of the lags are listed.}
\label{tab:lags}
\end{table}

The A-D test gives a significance of 16.8\,\% (the probability that the
two samples are drawn from the same distribution) for the short-intermediate
pair, and it yields a significance of 4.2\,\% for the long-intermediate pair.
Therefore, in case of  short and intermediate groups, we cannot reject the null
hypothesis that the two samples are drawn from the same distribution on a sufficiently
low level (5\,\%). On the other hand, this null hypothesis can be rejected in the case
of long and intermediate groups, but the significancy is not far below 5\,\% level.
The same test applied on the lags of the short-long pair yields a significance of
$<10^{-3}$\,\%. Therefore, in this case, the null hypothesis can be rejected
with a high significance. This strongly supports the well-known claim that the short and
long GRBs are really different phenomena and confirms the results of \citet{nor01}
(obtained with BATSE), but now by using the RHESSI instrument.

\subsection{Distribution of normalized lags}
\begin{figure}[h]
\centering
\includegraphics[trim=8mm 3mm 3mm 7mm,clip=true,width=0.48\textwidth]{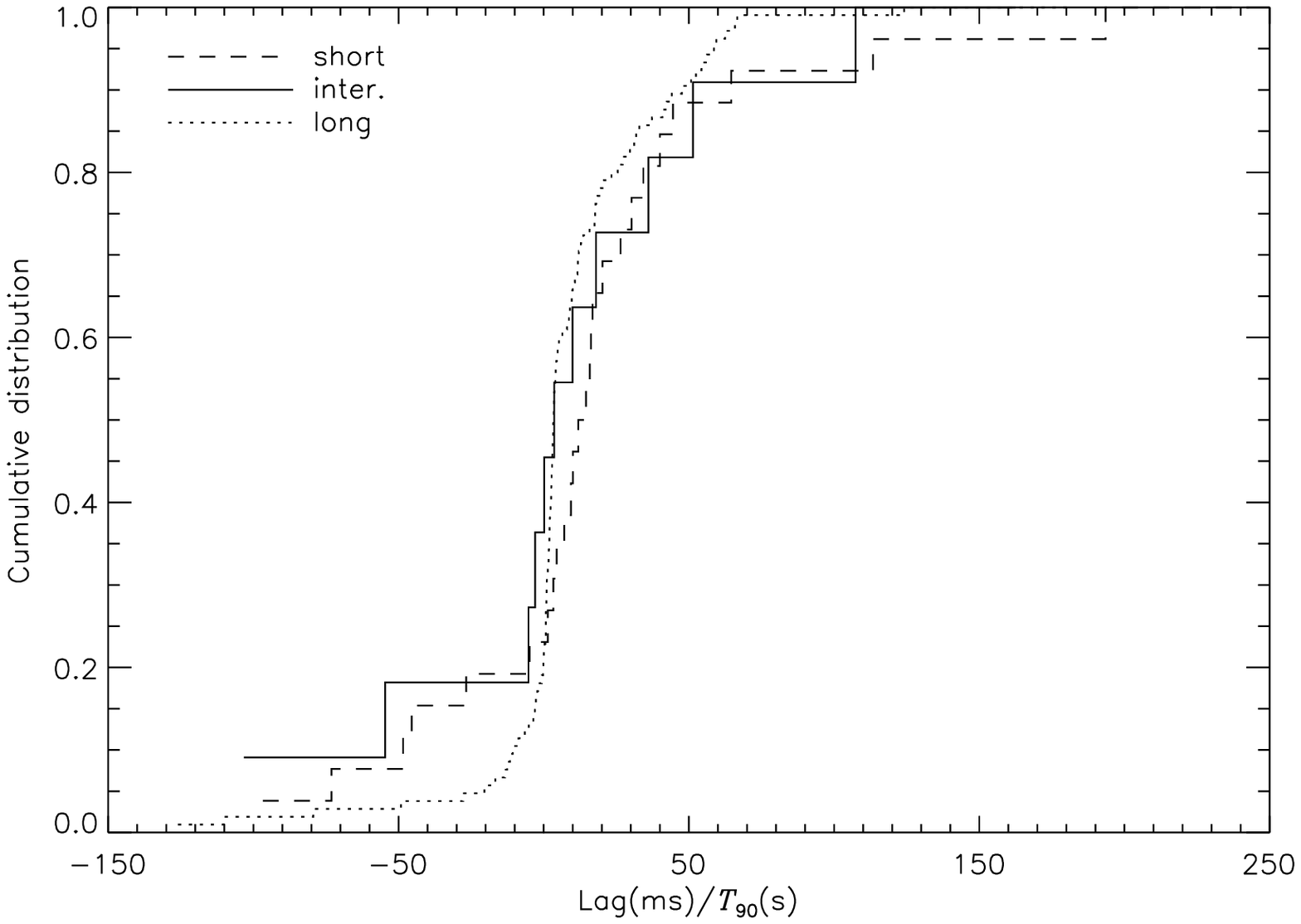}
\caption{The cumulative distributions of the normalized lags for the three RHESSI
GRB groups.}
\label{fig:normlags}
\end{figure}

In this section we compare the distributions of normalized lags (Fig.~\ref{fig:normlags}),
i.e. $L/T_{90}$, next to the absolute values of the lags. Again we use the A-D test between
the different GRB groups mentioned in the previous section. The number of events within the
groups is therefore the same. The results are summarized in Tab.~\ref{tab:normlags}.

The A-D test gives the significance level of 54.2\,\%
for the short-intermediate pair and it gives the significance
level of 45.0\,\% for the long-intermediate pair. The significances are considerably above 5\,\% level,
therefore the null hypothesis that the samples are drawn from the same distribution cannot be rejected.
For the short-long, pair the A-D test gives the significance level of 6.0\,\%.
If the normalized lags are concerned, the difference between the short and long bursts is not definite.

\begin{table}[h]
\centering
$
\begin{array}{cccc}
\begin{tabular}{ccc}
\hline\hline
Groups & A-D $P$ \\
       & (\%)    \\
\hline
Inter.-Short & 54.2 \\
Inter.-Long  & 45.0 \\
Short-Long   & 6.0 \\
\hline
\end{tabular}
\begin{tabular*}{1.0cm}{c}
\\
\end{tabular*}
\begin{tabular}{cccc}
\hline\hline
Group &  Mean                & Median              & $\sigma$ \\
      &  $L$(ms)/$T_{90}$(s) & $L$(ms)/$T_{90}$(s) &          \\
\hline
Short  &   21.3               &  15.8              & 63.3  \\
Inter. &   17.6               &  5.5               & 63.0  \\
Long   &   10.2               &  3.4               & 32.0  \\
\hline
\end{tabular}
\end{array}$
\caption{
Results of the A-D tests of the equality of the normalized lag
distributions between different RHESSI GRB groups are listed.
$P$ denotes the P-value of the test.
\emph{Right part:}
The means, medians and standard deviations $\sigma$ of the normalized lags are also mentioned.}
\label{tab:normlags}
\end{table}

\subsection{Distribution of peak-counts}

\begin{figure}[h]
\centering
$
\begin{array}{cc}
\includegraphics[trim=0mm 2mm 1mm 2mm,clip=true,width=0.48\textwidth]{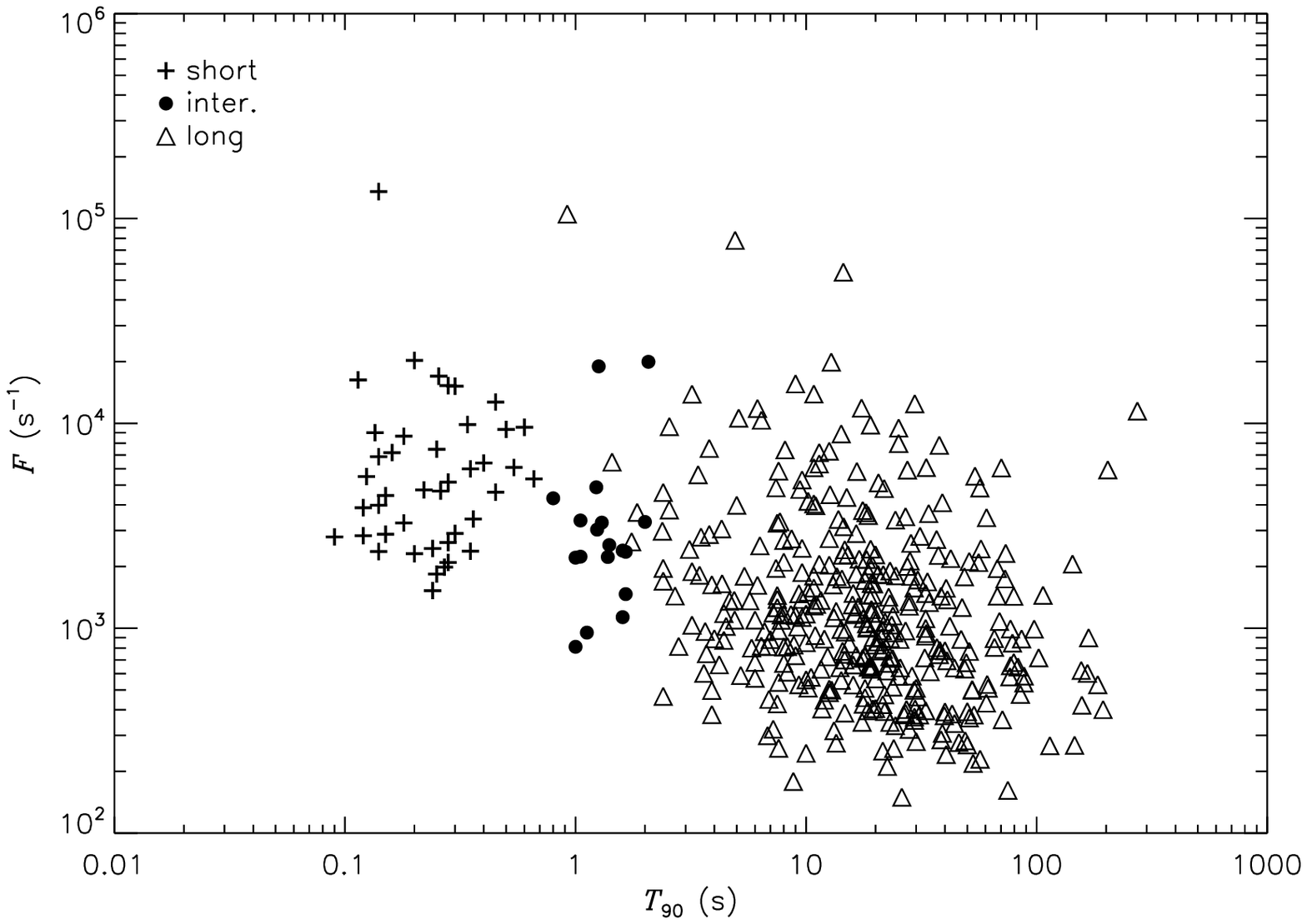}
&
\includegraphics[trim=0mm 2mm 1mm 2mm,clip=true,width=0.48\textwidth]{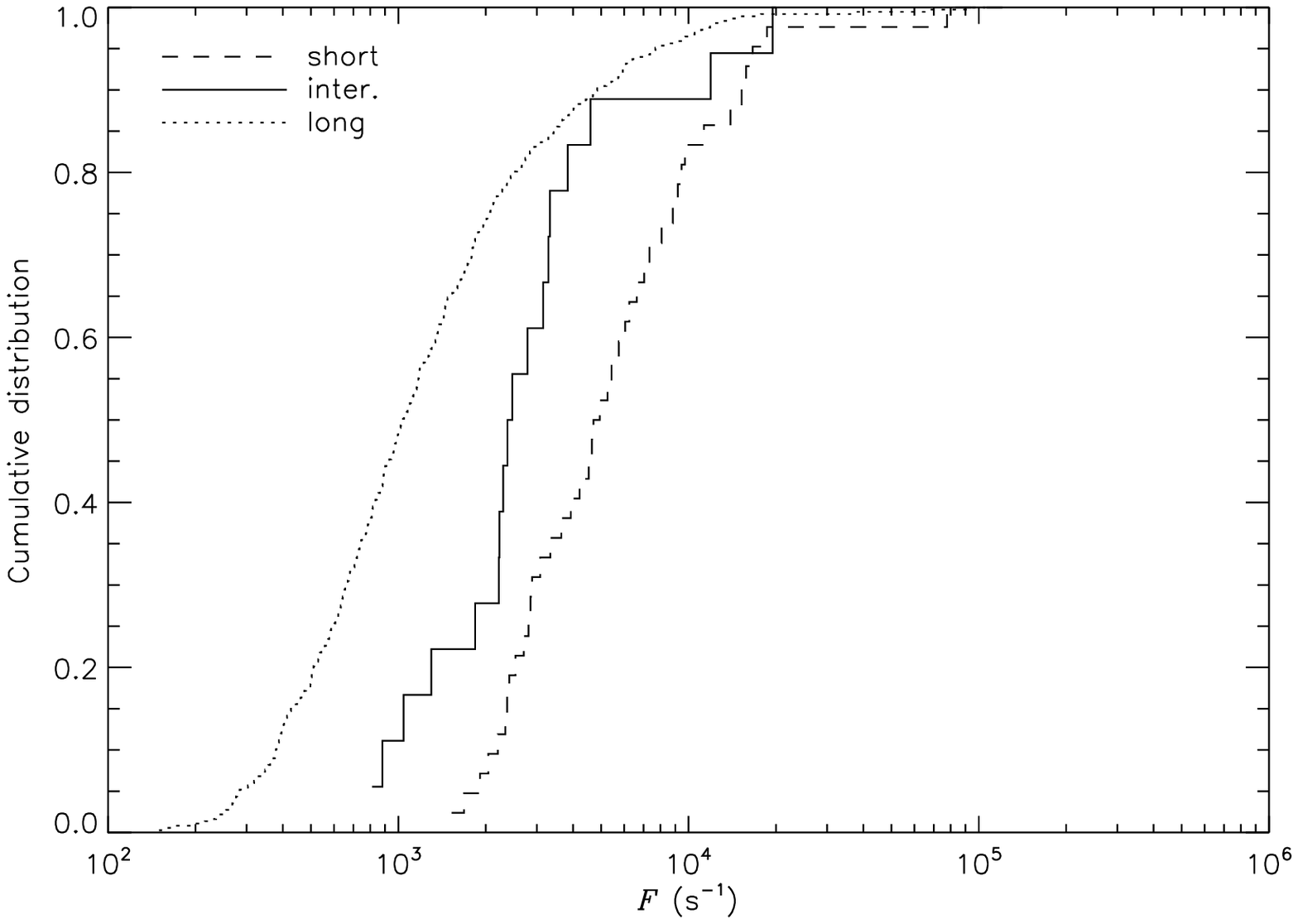}
\end{array}$
\caption{
\emph{Left panel:}
Peak-count rates $F$ of RHESSI GRBs as a function of $T_{90}$ durations for the three GRB groups,
identified by the analysis of the hardnesses and durations, are displayed.
\emph{Right panel:}
Cumulative distributions of these peak-count rates $F$ for the short-, intermediate-,
and long-duration bursts are shown.
}
\label{fig:peak_count_rates}
\end{figure}

Here we used Kolmogorov-Smirnov (K-S) test \citep{kol33,smir48} to compare the
cumulative distributions of the peak-counts among the different GRB groups.
The short (intermediate, long) group contains 42 (18, 367)
objects. The results are presented in Tab.~\ref{tab:peak_count_rates} and shown in
Fig.~\ref{fig:peak_count_rates}.

\begin{table}[h]
\centering
$
\begin{array}{cccc}
\begin{tabular}{ccc}
\hline\hline
Groups & $D$ & K-S $P$ \\
       &     & (\%)    \\
\hline
Inter.-Short &  0.44 & 0.9 \\
Inter.-Long  &  0.55 & 3$\times$10$^{-5}$ \\
Short-Long   &  0.69 & $<10^{-6}$ \\
\hline
\end{tabular}
\begin{tabular*}{1.0cm}{c}
\\
\end{tabular*}
\begin{tabular}{cccc}
\hline\hline
Group &         Mean        &       Median        & $\sigma$   \\
      &     $F$\,(s$^{-1}$) &     $F$\,(s$^{-1}$) & (s$^{-1}$) \\
\hline
Short &        9\,485       &        5\,163       &  20\,418  \\
Inter.&        4\,412       &        2\,546       &  5\,586   \\
Long  &        2\,589       &        1\,038       &  7\,673   \\
\hline
\end{tabular}
\end{array}$
\caption{
\emph{Left part:}
Results of the K-S test applied on the peak-count rates $F$ for the RHESSI database.
The K-S distance $D$ and the K-S significance $P$ are mentioned.
\emph{Right part:}
The means, medians, and standard deviations of the peak-count rates are listed.}
\label{tab:peak_count_rates}
\end{table}

The results of the K-S tests imply that the distributions of the
peak-count rates are different over all three groups. Particularly, the K-S significance
level for the intermediate vs. short busts is 0.9\,\%, for intermediate vs.
long bursts it is 3$\times$10$^{-5}\,\%$,
and for short vs. long bursts it is $<10^{-6}\,\%$.

\subsection{Monte Carlo simulations}
\label{sec:mc-sim}
In order to test the robustness of the results obtained by the A-D tests applied on
lags $L$, normalized lags $L/T_{90}$ and K-S tests applied on peak-count rates $F$,
one can use Monte Carlo method.

In case of spectral lags we proceeded in the following way: The procedure described in
Sec.~\ref{sec:sample} - calculation of statistical errors of the lags by applying of Poisson noise -
provided distribution of 1\,001 lags for each GRB. Thus for each GRB we randomly selected
one lag from its distribution and made 10\,000 data samples.
Then the A-D tests for these 10\,000 samples were calculated.

In case of peak rates, we proceeded as below: We applied the Poisson noise to the
measured light curves and subtracted the background in order to obtain the simulated data.
Then we derived the peak count rate for the same peak time when the peak was found in the
measured light curves. We proceed in this way for each GRB. Afterwards we calculated K-S tests and
repeated this sequence 10\,000 times.

\begin{figure}[h]
\centering
$
\begin{array}{cc}
\includegraphics[trim=9mm 3mm 1mm 6mm,clip=true,width=0.48\textwidth]{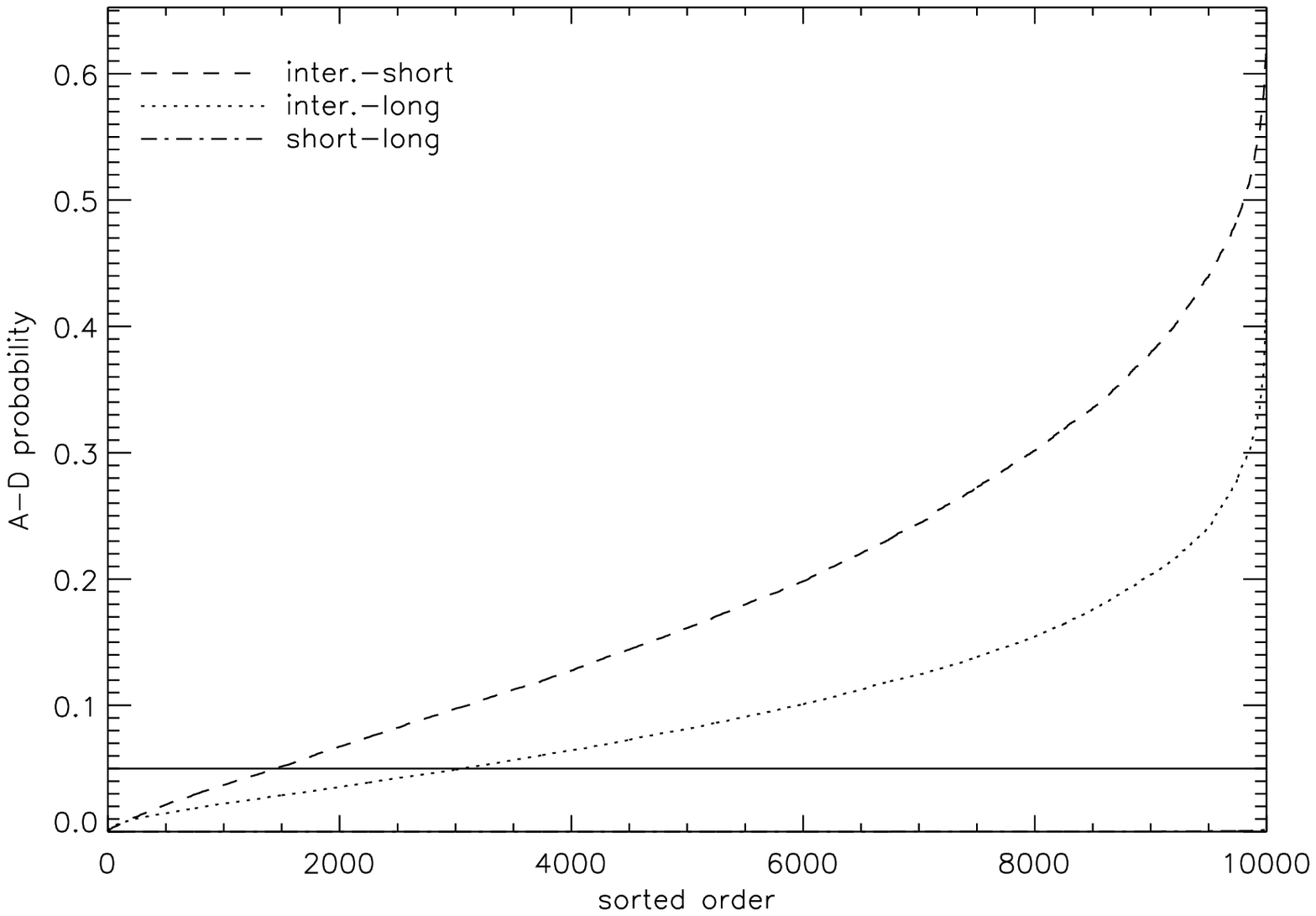}
&
\includegraphics[trim=9mm 3mm 1mm 6mm,clip=true,width=0.48\textwidth]{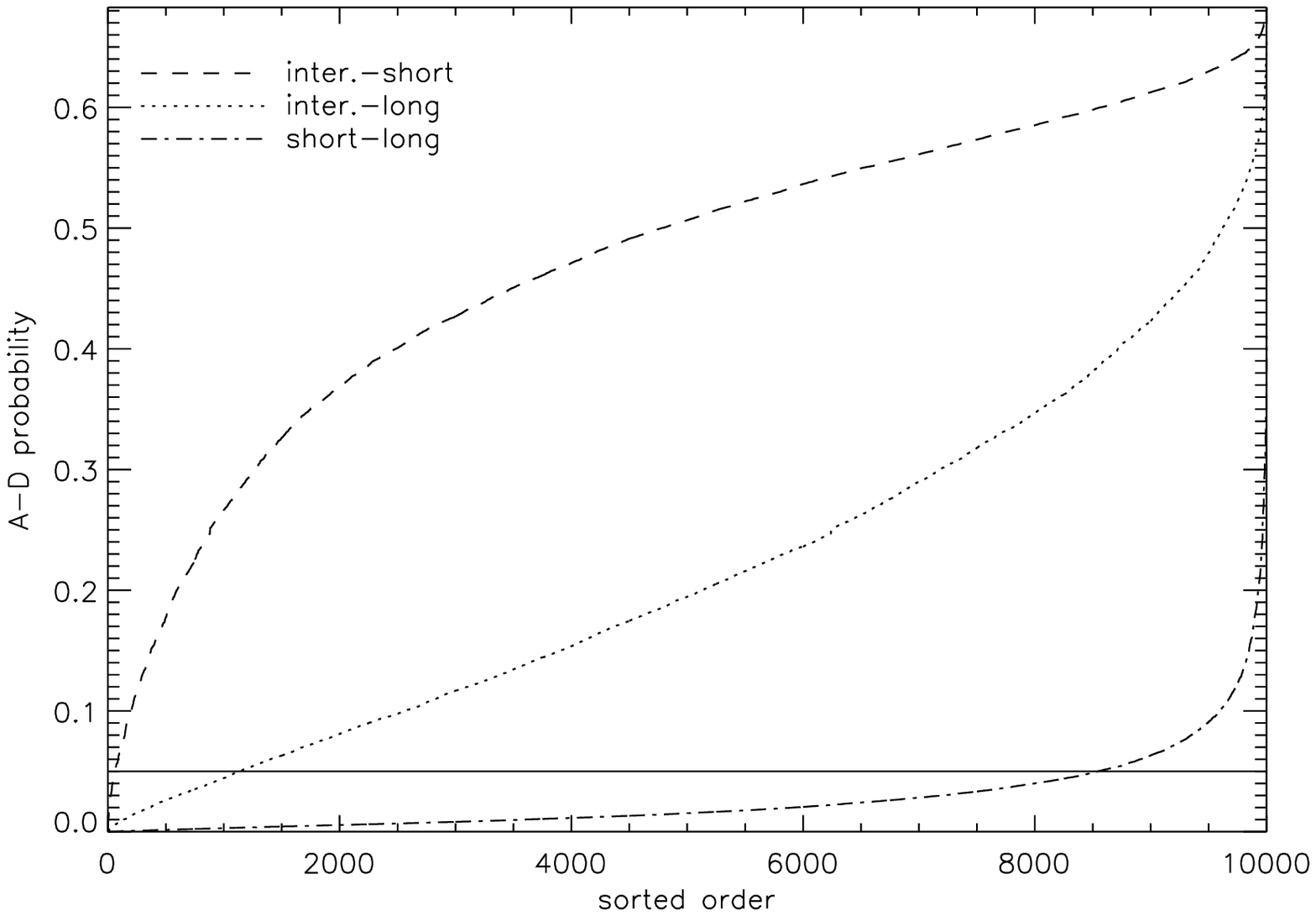}
\end{array}$
\caption{
\emph{Left panel:}
The A-D probabilities of the tests applied on the samples of lags (left panel) and normalized lags
(right panel) obtained from 10\,000 MC cycles for different GRB groups.
The horizontal solid line denotes the 5\,\% threshold.
}
\label{fig:lags_MC}
\end{figure}

The number of cases when the A-D and K-S probability reached higher values than 5\,\% for tests
done on different pairs of GRB groups is noted in Tab.~\ref{tab:montecarlo}. The results of
MC simulations comparing spectral lags and normalized lags are shown in Fig.~\ref{fig:lags_MC}.

The MC method confirms that the distributions of spectral lags between short
and long GRB groups are different. Let's compare results from the MC simulations of lags
and normalized lags between the intermediate-short and intermediate-long pairs
with the results of the tests applied directly on median lags (Tab.~\ref{tab:lags})
and median normalized lags (Tab.~\ref{tab:normlags}). Then one can see that MC simulations
gives A-D prob. $>$ 5\,\% more often then expected. It can be caused by the fact that for some
GRBs, weak and noisy ones, the distribution of lags found by MC method might not
follow the real distribution because after applying the Poisson noise the polynomial fit of
CCF may not well describe the CCF peak. The reason for this conjecture is that the fitting
range remained fixed and same for the simulated data as for the measured data. In other words
the suitable fitting range for the measured data need not be suitable for the simulated data.
In this case we think that the A-D tests applied on the median lags give more reliable results
than the MC simulations do. However, one mutual behaviour is seen here, the intermediate-short
pair has distributions of lags and normalized lags more similar than the pair intermediate-long does.
This feature is seen both in the A-D tests applied on the median lags/normalized
lags and in the A-D tests of the MC data samples.

\begin{figure}[h]
\centering
\includegraphics[trim=7mm 3mm 1mm 6mm,clip=true,width=0.48\textwidth]{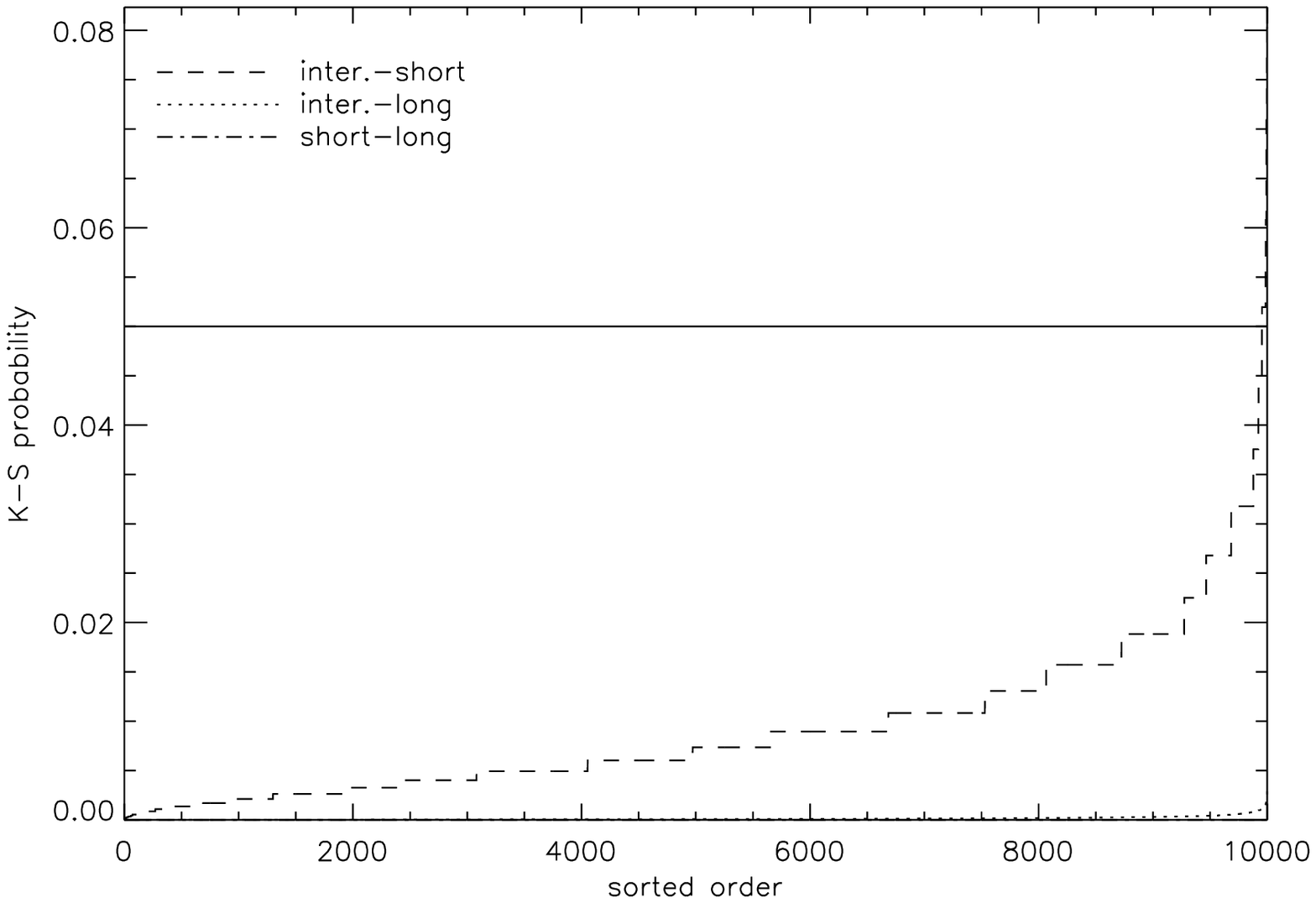}
\caption{The K-S probabilities of the tests applied on the samples of peak-count rates obtained from
10\,000 MC cycles for different GRB groups.}
\label{fig:F_MC}
\end{figure}

MC simulations also confirm results of K-S tests applied
directly on the measured peak rates. We can conclude that the short-,
intermediate- and long-duration bursts
have different distributions of peak count rates.
The results of MC simulations comparing spectral lags
and normalized lags are shown in Fig.~\ref{fig:F_MC}.

\begin{deluxetable}{cccc}
\tablecaption{Monte Carlo double-check of results from the statistical tests.
\label{tab:montecarlo}}
\tablewidth{0pt}
\tablehead{
\colhead{Tests} &
\colhead{Inter.-Short} &
\colhead{Inter.-Long} &
\colhead{Short-Long}
}
\startdata
Lags    		&  $8\,556$ ($85.6\,\%$) & $6\,938$ ($69.4\,\%$)        & \phn\phn\phd$0$ ($0.0\,\%$) \\
Norm. lags 		&  $9\,936$ ($99.4\,\%$) & $8\,862$ ($88.6\,\%$)        & $1\,458$ ($14.6\,\%$)     \\
Peak rates		&  \phn\phd$47$ ($0.5\,\%$) & \phn\phn\phn$0$ ($0.0\,\%$) & \phn\phn\phd$0$ ($0.0\,\%$) \\
\enddata
\tablecomments{The number of cases out of 10\,000 MC cycles (and their percentages)
are noted for A-D (lags and norm. lags) and K-S (peak rates) probability values exceeding 5\,\%
for tests done on spectral lags, normalized lags, peak-count rates, and on different pairs of GRB groups.}
\end{deluxetable}

\section{Discussion}
\label{sec:dis}

\subsection{Comparison with the BATSE database}

\begin{figure}[h]
\centering
$
\begin{array}{cc}
\includegraphics[trim=8mm 2mm 2mm
6mm,clip=true,width=0.48\textwidth]{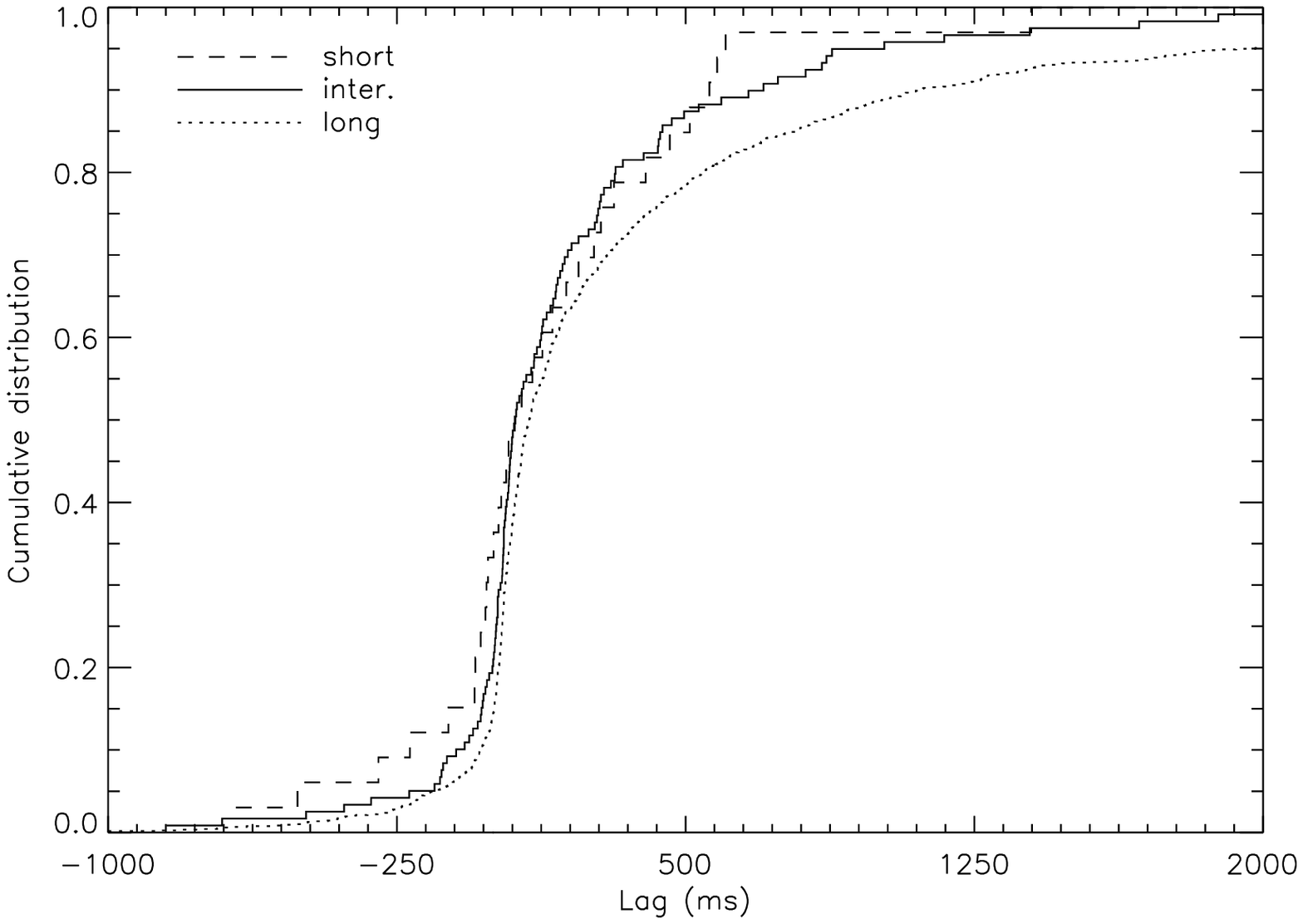}
&
\includegraphics[trim=8mm 2mm 1mm 5mm,clip=true,width=0.48\textwidth]
{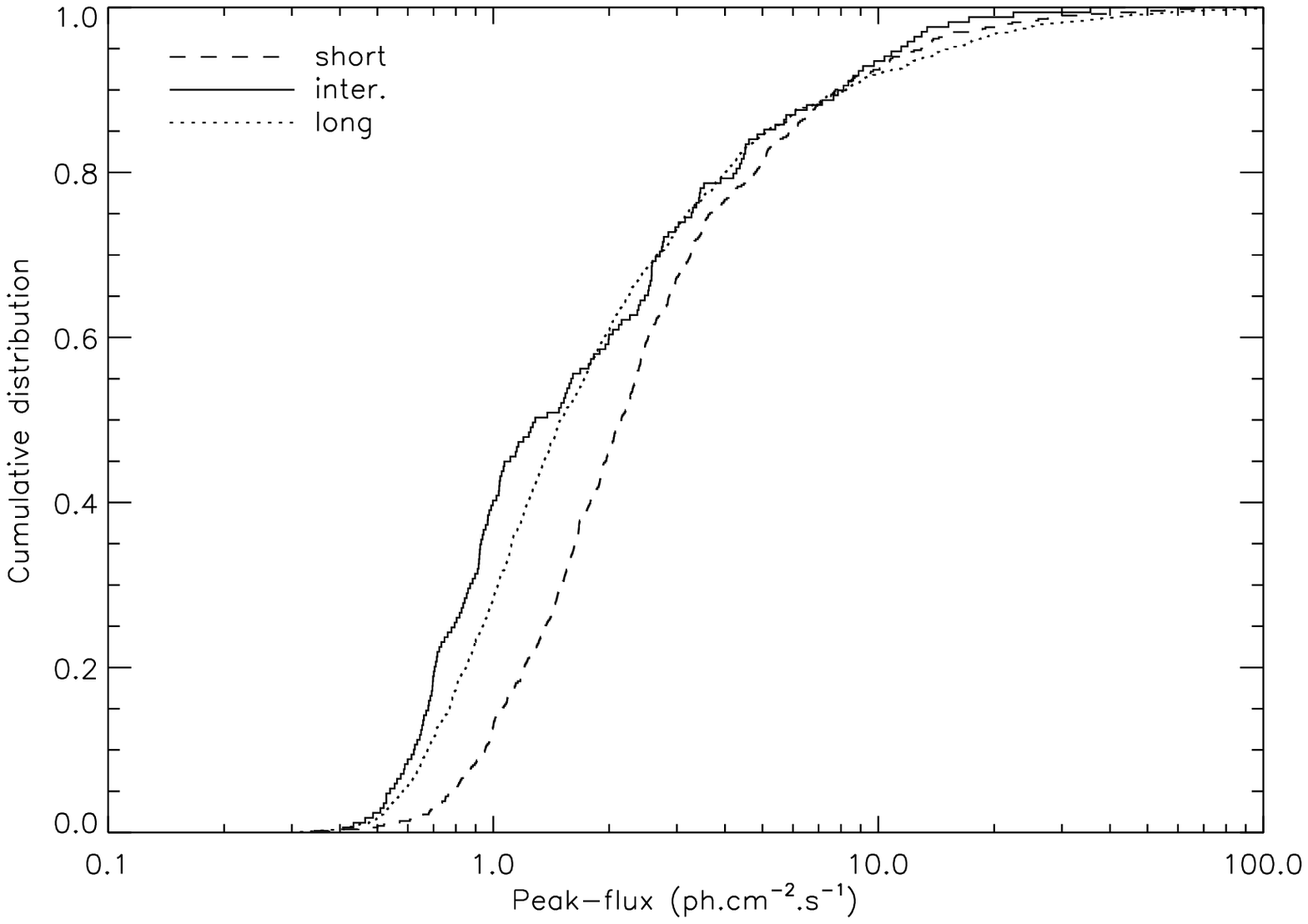}
\end{array}$
\caption{Cumulative distributions of the spectral lags (left panel)
and peak-fluxes (right panel) for the three BATSE GRB groups are shown.}
\label{fig:BATSE_lags}
\end{figure}

The lags of GRBs from the BATSE dataset are different for the short and long groups
\citep{nor01}: for the short bursts the lags on average are close to zero, but for
the long bursts they are positive. Norris and his collaborators did not study the
lags of the intermediate bursts separately \citep{nor01,nor02,nobo06}. For the sake
of completeness we have attempted to do this for the publicly
available data. \citet{ho06} defines membership within the groups for all BATSE GRBs.
Additionally, \citet{nor02} defines for any GRB with $T_{90} > 2\,$s its
lag\footnote{\url http://heasarc.gsfc.nasa.gov/docs/cgro/analysis/lags/web\_lags.html}.
Compilation of these two lists and the application of the A-D test on the lags of
the three BATSE groups (here for the first time)
produced the results collected in Tab.~\ref{tab:BATSE_lags} and shown in
Fig.~\ref{fig:BATSE_lags}. The short (intermediate, long) group contains 33 (119, 1179)
objects here. Of course, one must keep in mind that this sample is drastically truncated
for the short bursts. Hence, the short-intermediate and the short-long comparisons,
can serve only as a qualitative indicators. Even the intermediate-long pair cannot
be taken as representative because the truncation $T_{90} > 2\,$s can also omit
several intermediate GRBs.

\begin{table}[h]
\centering
$
\begin{array}{cccc}
\begin{tabular}{ccc}
\hline\hline
Groups       &  A-D $P$ \\
             &  (\%)    \\
\hline
inter.-short & 51.3 \\
inter.-long  & 3.8  \\
short-long   & 9.7  \\
\hline
\end{tabular}
\begin{tabular*}{1.0cm}{c}
\\
\end{tabular*}
\begin{tabular}{cccc}
\hline\hline
Group &  Mean $L$ & Median $L$ & $\sigma$ \\
      &  (ms)     &  (ms)      &   (ms)   \\
\hline
short  & 177.1  &  72.0 & 454.6 \\
inter. & 207.5  &  60.0 & 464.2 \\
long   & 390.7  &  94.0 & 848.2 \\
\hline
\end{tabular}
\end{array}$
\caption{
\emph{Left part:}
Results from the A-D test of the equality of the spectral
lag distributions for the BATSE GRBs. $P$ denotes the P-value of the test.
\emph{Right part:}
The mean, median and standard deviations $\sigma$ of the lags.}
\label{tab:BATSE_lags}
\end{table}

Keeping all this in mind, if the lags are taken into consideration,
we can say that there is some similarity between the BATSE and the
RHESSI databases. First of all, there is a similarity with regard to
the intermediate-long pair: the difference is confirmed,
though not on an high significance level, but remarkably the significances
from the A-D test are comparable (3.8\,\% and 4.2\,\%).
Second, there is a similarity with regard to the intermediate-short pair:
in both databases the A-D test reveals that for these two groups the distributions
of GRB lags are similar; the significances are 51.3\,\% for BATSE,
and 16.8\,\% for RHESSI. However, one must again keep in mind that our BATSE sample of
short bursts is truncated and the sample of intermediate bursts as well.
Third, both databases show a difference between the average lags for the short-long
pairs (for the BATSE databases the difference between the distributions
is shown not to be significant; the A-D P-value is only 9.7\,\%,
probably as a result of the sample truncation, but \citet{nor01} makes this claim unambiguously)

The results of the K-S tests applied on the peak-fluxes of the 64\,ms resolution light-curves
for the BATSE data imply that the distributions are different over all three groups.
The short (intermediate, long) group contains 502 (169, 1282) objects here, and the K-S tests are
summed in Tab.~\ref{tab:BATSE_peak_fluxes} and shown in Fig.~\ref{fig:BATSE_lags}.
These results are to be expected because, for example, \citet{na07} claims that the peak-fluxes
of short GRBs are roughly $20\times$ smaller than those of the long ones. It is also
known that the intermediate BATSE group is ``intermediate" concerning the fluence \citep{mu98}.

\begin{table}[h]
\centering
$
\begin{array}{cccc}
\begin{tabular}{ccc}
\hline\hline
Groups & $D$ & K-S $P$ \\
       &     & (\%)    \\
\hline
inter.-short &  0.30 & $<10^{-6}$ \\
inter.-long  &  0.13 & 1.0 \\
short-long   &  0.21 & $<10^{-6}$ \\
\hline
\end{tabular}
\begin{tabular*}{1.0cm}{c}
\\
\end{tabular*}
\begin{tabular}{cccc}
\hline\hline
Group &         Mean        &       Median        & $\sigma$ \\
      &          $F$        &          $F$        &          \\
\hline
short &        4.00         &         2.15        &   6.80   \\
inter.&        3.15         &         1.29        &   4.91   \\
long  &        4.09         &         1.51        &   10.31  \\
\hline
\end{tabular}
\end{array}$
\caption{
\emph{Left part:}
Results of the K-S test applied on the peak-fluxes $F$\,(ph.cm$^{-2}$.s$^{-1}$) of the
BATSE GRBs. The shortcuts have the same meaning as in Tab.~\ref{tab:peak_count_rates}.
\emph{Right part:}
The means, medians, and standard deviations of the peak-fluxes are listed.}
\label{tab:BATSE_peak_fluxes}
\end{table}

Therefore, our comparison of these RHESSI and BATSE groups finds similarities.
In the case of BATSE database, all three groups are different
in respect to two quantities (duration and peak-flux).
It is remarkable that for BATSE the hardness of intermediate group
is strongly anticorrelated with the duration \citep{ho06}. Since the hardness of the
intermediate group differs from the hardnesses of the short and long ones, these studies
support the opinion that all three BATSE groups represent different phenomena.

\subsection{Comparison with the Swift database}

The lags of the GRBs from the Swift dataset are also different for the short and long groups
\citep{ugar11}. \citet{ugar11} also discuss the lags of the intermediate bursts,
and they find a behavior which does not resemble the cases found in
the RHESSI and BATSE datasets.
The Swift's intermediate-long pair has on average similar lags, but there is a statistically
significant difference in the short-intermediate pair. Thus, if the lags are considered,
the Swift's intermediate group is similar to the Swift's long group \citep{ugar11}.
On the other hand, the peak-fluxes differ significantly in the short-intermediate and intermediate-long pairs,
respectively. The peak-fluxes of the short-long pair are not different from the statistical point of
view \citep{ve10}. Nevertheless, \citet{ugar11} concludes that ``Swift's intermediate
bursts differ from short bursts, but exhibit no significant differences from long bursts apart from
their lower brightness". In other words, in the Swift database there is a clear similarity between
the intermediate group and the long one. The physical difference of the short
and long bursts in the Swift database further holds \citep{ve10,ugar11}.

Comparison with the Swift's groups leads to the conclusion that
the third group in the Swift database is strongly related to the long group, as stated by
\citet{ve10} and \citet{ugar11}, and only the short group should represent another phenomenon.
There is a difference in the hardness, peak-flux, and duration for the intermediate-long pair
\citep{ho08,ve10}, but no clear separation occurs for the lags \citep{ugar11}.
We have no reason to query the conclusions of \citet{ve10}
that the intermediate group is related to XRFs which in turn can be
related to standard long GRBs. We add that the separation within
the long group itself into harder and softer parts is not fully new \citep{pe97,tav98}.
We allow to claim that the intermediate-duration bursts in the RHESSI and Swift
databases are different phenomena. These results followed exclusively from the statistical analyses.

\subsection{Discussion of the number of groups}
\label{sec:H-T90}
In order to provide an extended discussion on the number of GRB groups
we apply clustering methods to our data sample.
This also serves to extend the statistical analysis performed by \citet{rip09}.
In general, the clustering methods can be divided into
parametric and non-parametric types. Parametric methods assume that the data follow a pre-defined model
(in our case a sum of multivariate Gaussian functions). These methods assign for each GRB a
probability of membership in a certain group. The non-parametric methods, e.g. K-means clustering,
provide definite assignments of each burst to a given group.
More details about these methods can be found in the book by \citet{eve11}.
Model-based clustering is also described in \citet{mcla00}.

We apply model-based clustering and K-means clustering methods on
our RHESSI data sample by using the algorithms
implemented in the R software.

\subsubsection{Model-based clustering method}
\label{sec:model-based-clust}
In this method we assume that the distribution of the tested parameters
(logarithms of durations, hardness ratios, peak-count rates, and normalized lags)
follow a superposition of Gaussian functions.
Similar analysis for GRB classification was done by \citet{mu98,ho06} and \citet{ve10}.

The Maximum Likelihood method is used to find the best-fitted model parameters.
Adding more free parameters to a fitted model can increase the likelihood, but also may result in
overfitting. It is possible to penalize a model for more free parameters.
This can be done by a method called the Bayesian Information Criterion
(BIC) presented by \citet{schw78}. The function which must be maximized to get the
best-fitted model parameters is:
BIC = 2\,ln\,l$_\mathrm{max}$ - $m$\,ln\,$N$, where l$_\mathrm{max}$ is the maximum
likelihood of the model, $m$ is the number of free parameters, and
$N$ is the size of the sample. In our work we use the BIC to determine the most probable model,
its parameters and the number of its components.

For model-based clustering, we use \emph{Mclust}
package\footnote{http://cran.r-project.org/web/packages/mclust/index.html}
\citep{fra00} of R. For the explanation of different models, see the Mclust
manual\footnote{http://www.stat.washington.edu/research/reports/2006/tr504.pdf}.
The nomenclature of the different models in Mclust involves the following designations:
the volumes, the shapes and the orientation of the axes of all clusters may be equivalent (E) or may
vary (V) and the axes of all clusters may be restricted to parallel orientations
with the coordinate axes (I).

\subsubsection{Model-based clustering - 2 variables}
First, we start with a two-dimensional case and fit $T_{90}$ durations and hardnesses $H$.
The data sample consists of 427 bursts (Table~7 of \citet{rip09} and Table~\ref{tab:dec-corr}).

In this case the number of free parameters of the model with $k$ bivariate
Gaussian components is $6k - 1$ ($2k$ means, $2k$ standard deviations, $k$ correlation coefficients,
and $k-1$ weights, because the sum of the weights is 1). For the most general model, all parameters
are free. However, sometimes we want to test models where some of the parameters
between different components are in a relation with other parameters, e.g. all components
have the same weight or shape, etc. In this case the number of degrees of freedom
is reduced.

As seen in Fig.~\ref{fig:BIC-T90H} the best fitted model has $k=2$ components with equal volumes,
variable shapes, and with the axes of all clusters parallel with the coordinate axes (EVI model).
This best fitted model has a value of $\mathrm{BIC} = -681.5$. The EVI model with one component
gives $\mathrm{BIC} = -899.1$ and with three components $\mathrm{BIC} = -701.8$.
For all other tested models with $k=1$ component the highest BIC is -820.3 and
with $k=3$ components -694.3, which are clearly below the maximum.

The difference between the BIC of two models gives us information
about the goodness. According to \citet{kas95} and \citet{mu98}, a difference in BIC of
less than 2 represents weak evidence, difference between 2 and 6 represents positive evidence, between
6 and 10 strong evidence, and difference greater than 10 represents very strong evidence in favor of
the model with the higher BIC.

In our case the difference between the best fitted model (EVI) with two components and
the EVI models with one or three components is always higher than 10.
This gives a strong support for the EVI model with $k=2$ components.

The two components are short/hard and long/soft groups. The intermediate-duration bursts
showed in Fig.~\ref{fig:groups} are assigned to the short/hard group by this test.

\subsubsection{Model-based clustering - 3 variables}
\label{sec:H-T90-F}
Next we perform model-based clustering on three variables:
$T_{90}$ durations, hardnesses $H$, and peak-count rates $F$.
Since the peak rates were measured for all events,
the sample here also consists of all 427 bursts (Table~\ref{tab:database}).

\begin{figure}[h]
\centering
$
\begin{array}{cc}
\includegraphics[trim=31mm 2mm 9mm 16mm,clip=true,height=0.47\textwidth,angle=-90]{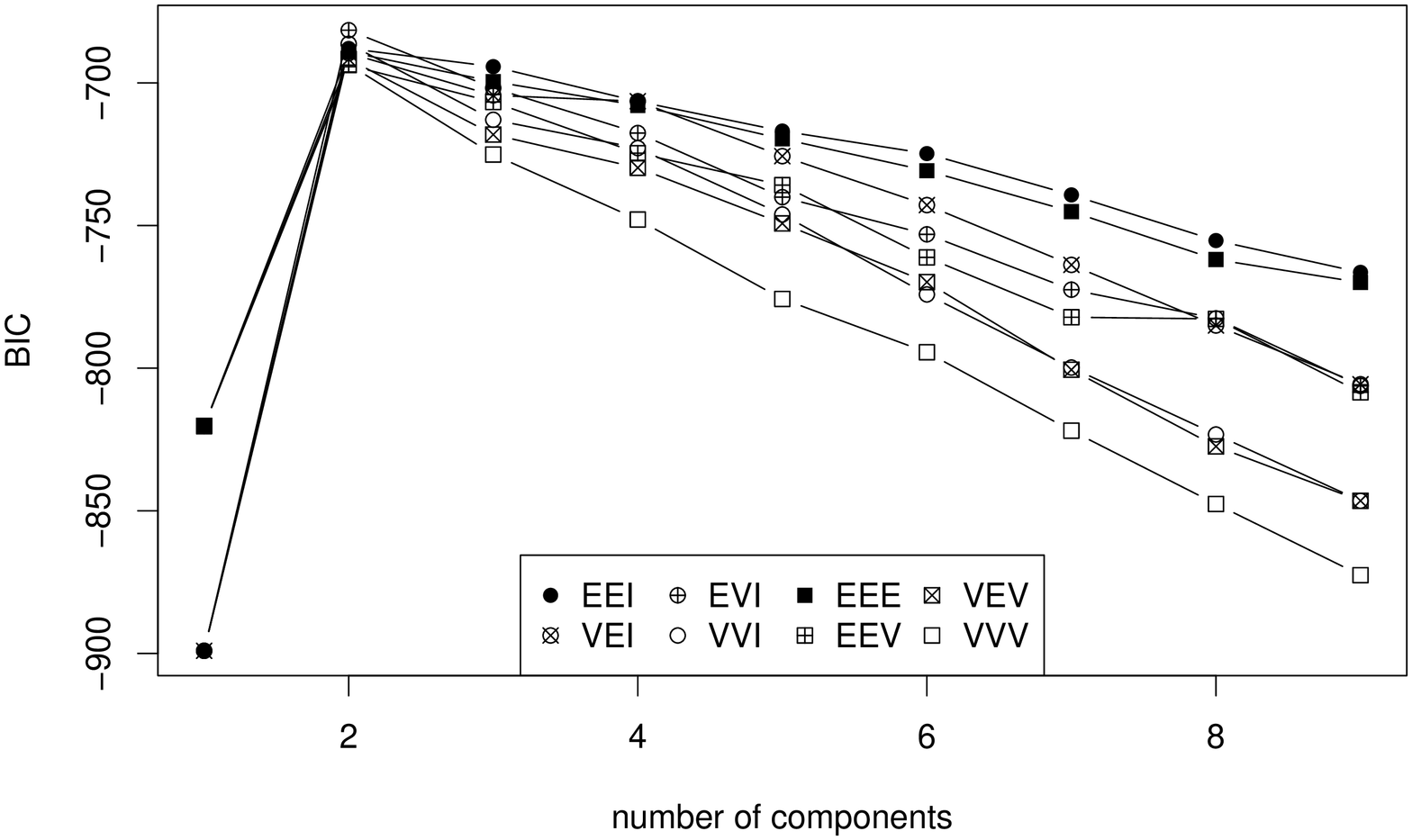}
&
\includegraphics[trim=31mm 2mm 9mm 16mm,clip=true,height=0.47\textwidth,angle=-90]{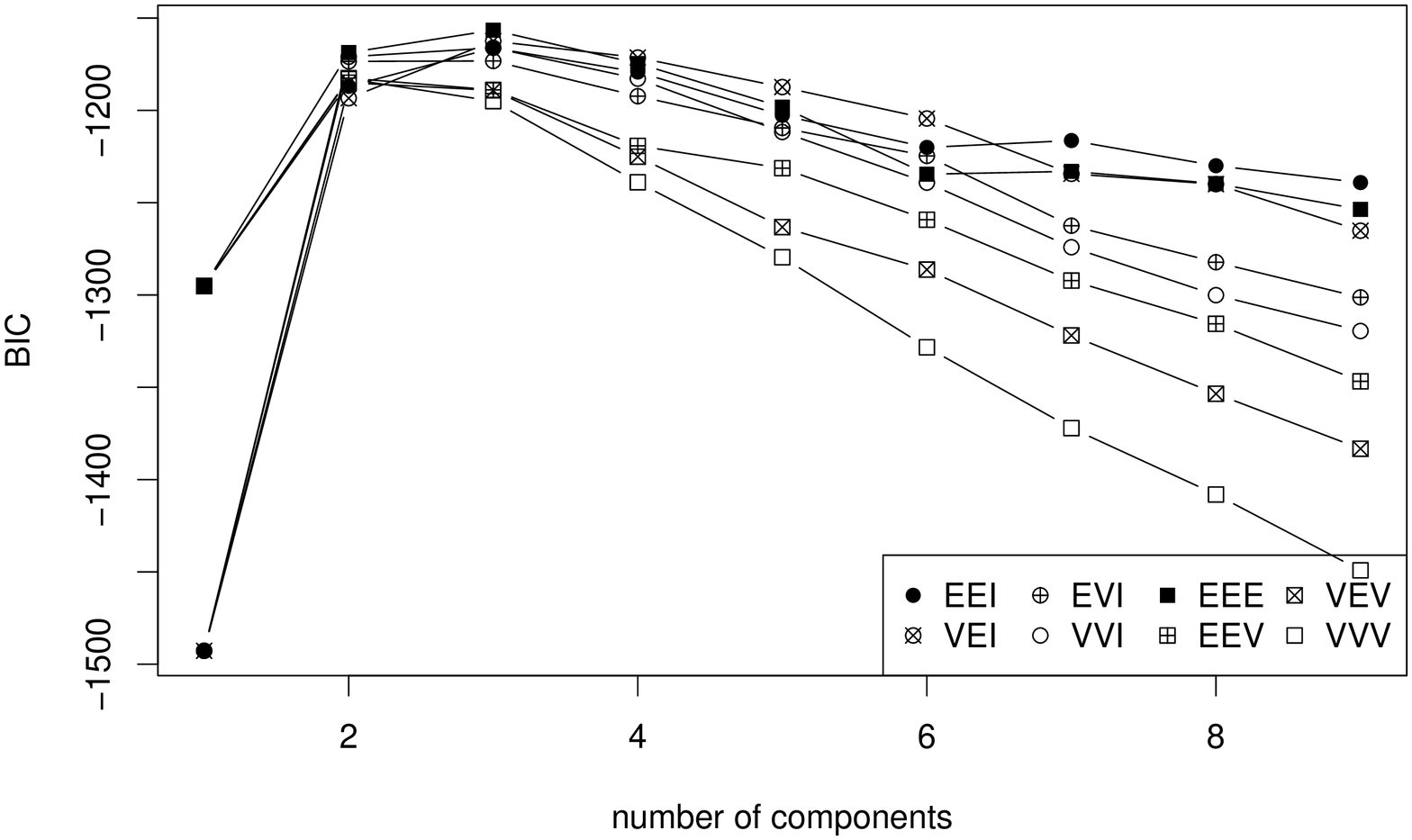}
\end{array}$
\caption{
\emph{Left panel:}
Bayesian information criterion (BIC) values for different models as a function of the number
of bivariate Gaussian components. The higher the BIC value, the more probable the model.
The most probable model is EVI with two components. The data sample consists of two variables:
$T_{90}$ durations and hardness ratios.
\emph{Right panel:}
BIC values for different models plotted against the number of components.
The most probable model is EEE with three components.
The data sample consists of three variables: $T_{90}$ durations,
hardness ratios $H$, and peak-count rates $F$.}
\label{fig:BIC-T90H}
\end{figure}

The best fitted model has $k=3$ components (see Fig.~\ref{fig:BIC-T90H})
with equal volumes, equal shapes and equal correlation coefficients between all clusters (EEE model).
This best model has a value of $\mathrm{BIC} = -1156.6$.
The EEE model with two components gives $\mathrm{BIC} = -1168.7$ and
for four components $\mathrm{BIC} = -1174.6$.
Markedly high values of BIC are also obtained for the EEI, VEI, and VVI models
with $k=3$ components, $\mathrm{BIC} = -1166.2$, $\mathrm{BIC} = -1162.5$, and
$\mathrm{BIC} = -1166.1$, respectively.

\begin{figure}[h]
\centering
\includegraphics[trim=9mm 9mm 9mm 9mm,clip=true,height=0.9\textwidth,angle=-90]{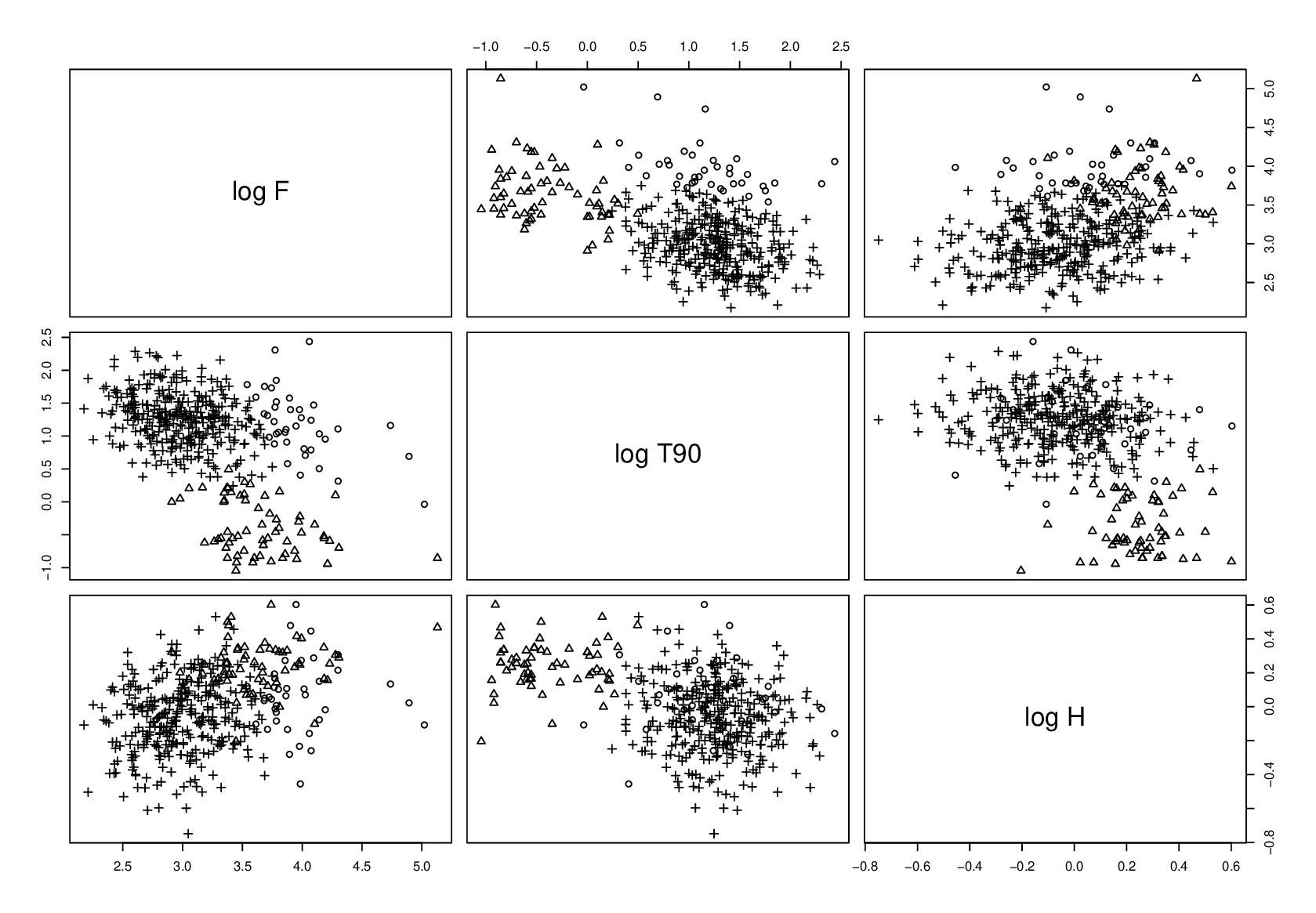}
\caption{A scatter plot of 427 bursts, with measured $T_{90}$, $H$, and $F$ assigned into three
groups by the EEE model.}
\label{fig:BIC-FT90H-scattplot}
\end{figure}

The difference in BIC between the EEE model with three components and
the EEE models with two or four components is $>10$.
The other models with other number of components (except above-mentioned EEI, VEI,
and VVI models with three components) gives BIC value lower by at least 10.
This provides a strong evidence in the favor of EEE
model with $k=3$ components. The group structure of this model with three
components is shown in Fig.~\ref{fig:BIC-FT90H-scattplot}.

The intermediate-duration bursts showed in Fig.~\ref{fig:groups} are assigned to the
short/hard group by this test. A new result here is that the group of long bursts is separated
into high- and low-peak flux clusters.

\subsubsection{Model-based clustering - 4 variables}
\label{sec:H-T90-F-L/T90}
Spectral lags of BATSE GRBs, i.e. the time delay between low and high energy photons from short and long
groups have been found to differ. For short bursts, an average
lag is $\sim 20-40$ times shorter than for long bursts, and the lag distribution is close to
symmetric about zero - unlike long bursts \citep{nor01,nor02,nobo06}. This result gave us
an idea to incorporate the spectral lags as well.

In this part we apply the model-based clustering on GRB peak-count rates $F$,
$T_{90}$ durations, hardness ratios $H$, and as a new addition to the variables, on normalized
lags $L/T_{90}$. Since the RHESSI spectral lags were calculated only for 142 bursts
(Table~\ref{tab:database}), our sample is truncated.

The best fitted model has $k=2$ components and is unconstrained, i.e. it has variable volumes,
variable shapes and variable correlation coefficients (VVV).
The best BIC value for this model is $\mathrm{BIC} = -1768.4$.
However, the VVV model with $k=3$ components gives similar value of $\mathrm{BIC} = -1768.5$.
The other models gives BIC value lower by at least 10. This strongly supports the VVV model with
$k=2$ components. There is no need to introduce VVV model with three components that has more free parameters.
The two components are separated accordingly to the values of normalized lags into zero- and non-zero-lag events.

\subsubsection{Summary of model-based clustering}
The model-based clustering of two-parameter data ($T_{90}$ and $H$) gives a strong evidence
in favor of EVI model with two components. The analysis of three-parameter data ($T_{90}$, $H$ and $F$)
shows that the best fitted model is EEE with three components. Surprisingly a new result is obtained here;
the group of long bursts is separated into high- and low-peak flux clusters.
The analysis of four-parameter data ($T_{90}$, $H$, $F$ and $L/T_{90}$) supports
the VVV model with two components only. The separation into the two components here is accordingly
to the values of normalized lags into zero- and non-zero-lag events.

\begin{deluxetable}{ccccccccc}
\tabletypesize{\footnotesize}
\tablecaption{A summary of the results from the model-based clustering.
\label{tab:discussion-tests}}
\tablewidth{0pt}
\tablehead{
\colhead{} &
\colhead{Model} &
\colhead{$k$} &
\colhead{$\mathrm{BIC}$} &
\colhead{$\mathrm{\Delta BIC}$} &
\colhead{$\mathrm{\Delta BIC}$} &
\colhead{$\mathrm{\Delta BIC}$} &
\colhead{$\mathrm{\Delta BIC}$} &
\colhead{Evidence} \\
\colhead{}&\colhead{}&\colhead{}&\colhead{}&\colhead{k=1}&\colhead{k=2}&\colhead{k=3}&\colhead{k=4}&\colhead{}
}
\startdata
2 par. & EVI   & 2         & -681.5           & $>10$ & $\times$      & $>10$    &                & very strong \\[0.5ex]
\hline\\[-2ex]
3 par. & EEE   & 3 & -1156.6 &       &$>10$ & $\times$ & $>10$ & very strong \\[0.5ex]
\hline\\[-2ex]
4 par. & VVV   & 2 & -1768.4 & $>10$ & $\times$      & $>10$    &                & very strong \\[0.5ex]

\enddata
\tablecomments{The results for model-based clustering applied on 2, 3, and 4 parameters is presented.
The values of BIC for the best fitted models with $k$ components are listed, as well as the
differences to the models with other number of components.}
\end{deluxetable}

\subsubsection{K-means clustering}
\label{K-means-HT90}
One of the non-parametric clustering methods is K-means \citep{mac67}.
Before we use our data for this method we scale them, i.e. we subtract
the mean value and then divide them by the standard deviation.
The reason for this procedure is that the clustering algorithm is sensitive to the distance
scale of the variables. For more details about the application of the K-means
method in a similar analysis of GRB data, see, e.g. \citet{cha07} or \citet{ve10}.
For this clustering method we use \emph{kmeans} package implemented in the R software.

To use the K-means method, one must set the number of clusters beforehand.
Then the corresponding number of centers is found by minimizing the sum of
squared distances from each burst to the center of the group to which they belong.
There is no precise way to determine the best number of clusters
with this method. However, it has been suggested that if one plots the within-group
sum of squares (WSS) as a function of the number of clusters, then an ``elbow" will indicate
the best number \citep{har75}. This method do not provide any probability indicating the
significance or insignificance for the given best number of clusters.

The calculated WSS as a function of the number of groups for our data samples using 2 ($T_{90}$, $H$),
3 ($T_{90}$, $H$, $F$), and 4 ($T_{90}$, $H$, $F$, $L/T_{90}$) variables are rather smooth and do not
demonstrate any remarkable and sharp ``elbows'' and thus do not bring useful information on the GRB classification.

\subsection{Discussion of the results}

The K-S tests applied on peak-count rates show that the distributions
are different over all three groups. The K-S significance level
for the short-long pair is $<10^{-6}$\,\%,
for the intermediate-long pair it is $3\times$10$^{-5}$\,\%,
and for the intermediate-short one it is 0.9\,\%.
The short and long GRBs have clearly different distributions of peak rates.
Also the intermediate and long GRBs have clearly different distributions of peak rates.
The intermediate-short pair also exhibits different distributions (K-S probability $<5$\,\%),
however less markedly than the other pairs of groups do.
These results are confirmed by MC simulations.

The A-D tests applied on distributions of spectral lags unveil that
the A-D probability for the short-long pair is $<10^{-3}$\,\%,
for the intermediate-long pair the A-D probability is 4.2\,\%,
and for the intermediate-short one it is 16.8\,\%.
The short and long GRBs have clearly different distributions of spectral lags.
The intermediate and long GRBs have A-D probability $<5$\,\%, however in this case the difference is not strong.
The intermediate-short pair does not exhibit different distributions.
The difference in the spectral lag distributions of the short-long pair of GRB groups is confirmed by MC simulations.
In case of intermediate-short and intermediate-long pairs the MC simulations reveal the same tendency as the A-D tests
applied directly on the measured values, i.e. the intermediate-short pair has more similar
distributions of spectral lags than the intermediate-long pair has. However, MC simulations gives A-D probability
higher than 5\,\% more often then expected. A possible reason is commented in Sec.~\ref{sec:mc-sim}.

The A-D tests applied on distributions of normalized lags show that
these distributions cannot be claimed as different. The A-D probability for the short-long pair is 6.0\,\%,
for the intermediate-long pair it is 45.0\,\%, and for the intermediate-short one it is 54.2\,\%.
Here one can see the same tendency as in the case of A-D tests applied on spectral lags, i.e.
the short-long pair has the least similar distributions of lags, the intermediate-long couple stays in the middle,
and the intermediate-short pair has the most similar distributions.
This tendency appears also in the MC simulations, however the absolute frequency of the cases when A-D probability
exceeds 5\,\% level happens more often then expected. The reason could be the same as in the case of MC simulations
applied on absolute values of spectral lags.

The model-based clustering of two-parameter data ($T_{90}$ and $H$) gives a strong evidence
in favor of a model with two components only. The two components are short/hard and long/soft groups.
The intermediate-duration bursts showed in Fig.~\ref{fig:groups} are assigned to the short/hard group by this test.
The analysis of three-parameter data ($T_{90}$, $H$ and $F$) shows that the best fitted model has three components.
Surprising point here is that this method separates the group of long bursts into high- and low-peak flux clusters.
The analysis of four-parameter data ($T_{90}$, $H$, $F$ and $L/T_{90}$) supports a model with two components which
are separated accordingly to the values of normalized lags into zero- and non-zero-lag events.

Surveying the A-D and K-S tests of the RHESSI data, it should be raised
that the difference between the short and long bursts was again strongly confirmed.
This follows from the different distributions of the spectral lags
and by the different distributions of the peak-count rates, both results were confirmed by the MC method.
This is already an expectable result, but - usefully - this result came from a new observational database.

According to Fig.~\ref{fig:groups} the intermediate-short pair of groups have similar hardness ratios.
Also according to the results of the A-D test of the spectral lags, the distributions of lags are not different
for the intermediate-short pair. However, the intermediate-duration and short-duration bursts are not completely same
because their peak-count rate distributions differ. On the other hand, the intermediate-long pair of groups differs
in hardness ratios, spec. lags, and peak-count rates.
Therefore, in our opinion, it is possible that the intermediate group detected by RHESSI
in Sec.~\ref{sec:sample} and by \citep{rip09} may be a longer tail of standard short/hard bursts.
This can be supported also by the fact that model-based clustering method applied on hardness ratios and durations
unveil only two clusters as the best solution; classical short/hard and long/soft groups, the intermediate-duration
bursts are assigned to the short group.

The RHESSI intermediate and long groups seem to be different phenomena.
This difference is supported by the distribution of the peak-count rates and spectral lags.
The results show that the intermediate group is also ``intermediate" with regard to its lags.
The intermediate group detected by Swift was found to be related to XRFs \citep{ve10},
and those may in turn belong to the standard long GRBs \citep{kip03}.
In the case of RHESSI, the longer and softer GRBs are
more difficult to be detected, because RHESSI's sensitivity declines rapidly below
$\approx50$\,keV and the weak and soft GRBs are not easily observable \citep{rip09}.
On the other hand, Swift is less sensitive in the photon-energy range $>150$\,keV.
But softer GRBs are readily detectable with this instrument.
Hence, in our opinion, an instrumental effect may be responsible for that the two satellites
(Swift and RHESSI) detected different intermediate groups.
This means that - from the statistical point of view - different groups
can be found if one looks at different databases.

There are bursts observed with properties similar to the short bursts
(hardness, lag) except their durations exceed 2\,s. For example, beyond \citet{ge06} and
\citet{kan11} mentioned already in Introduction,
\citet{nobo06} claims that ``short bursts with extended emission
(SGRBEE) can have $T_{90} > 2$\,s". Furthermore, also others \citep{deba11}
propose the astrophysical fragmentation of the short GRB group.

Concerning SGRBEE, we inspected the light-curves of all 18 RHESSI intermediate bursts,
but we found no softer extended emission coming after the main hard spike as is
typical for this kind of bursts. Figure~3 of \citet{per09} shows that the average $T_{90}$ duration
of the initial spike of an SGRBEE lies between the average durations of short and long bursts.
If RHESSI detects only the hard initial spike, and the softer extended emission is lost
in the noise, then the detected intermediate group might be polluted
by these objects.
Therefore, we also checked the light-curves of the RHESSI intermediate bursts as observed
by Konus-Wind \citep{apt95}, because it also has a good sensitivity below 50\,keV
(its range is $10-10\,000$\,keV). It has an overlap with the following RHESSI intermediate bursts:
GRB~020819A, GRB~030410, GRB~040329, GRB~050530, GRB~070802, GRB~070824, GRB~080408.
However, no extended emission was observed for these seven bursts by Konus-Wind.
This observation indicates the RHESSI intermediate GRBs
should not be dominantly polluted by SGRBEEs.

Furthermore, there are also additional indications that GRBs
which do not belong to the long+XRF pair category,
may originate from a broad range of astrophysical phenomena. For example, \citet{mu98} found four
subclasses in the BATSE database from the year 1998, but the fourth group was populated by a single
GRB. From the statistical point of view, such an object is an outlier of uncertain origin.
Likewise, similar situations exist concerning the objects GRB~060614 \citep{ge06} and
GRB~110328A \citep{cu11}. Any study of such a single unusual object is beyond
the scope of this article providing only statistical analyses.

\section{Conclusions}
\label{sec:sum}

The main results of this study can be summarized as follows:

\begin{itemize}
\item
Maximum Likelihood test in the duration-hardness plane of 427 RHESSI GRBs,
taken from \citet{rip09} but now with six events corrected for decimation,
again exhibits statistically significant third, in duration intermediate, group.
This completes the work \citet{rip09} using the durations and hardnesses only.

\item
The spectral lags and peak-count rates have been calculated for GRBs
observed by the RHESSI satellite for the first time.
The spectral lags were obtained for 142 objects, and the peak-counts for all 427  GRBs.
Hence, we constructed a new observational database for this satellite.
Then the three GRB subgroups were analyzed statistically
with respect to these new spectral lags and peak-count rates.

\item
The difference between short and long groups has been confirmed.
Usefully this result came from a new observational database.

\item
Kolmogorov-Smirnov and Anderson-Darling tests applied on spectral lags and peak-count rates
indicate that the intermediate group in the RHESSI database might be a
longer tail of the short group or at least has some common properties with this short group.
Contrary to this, the intermediate and the long groups are different.

\item
The group of RHESSI intermediate-duration GRBs is not dominantly populated by SGRBEEs.

\item
The intermediate-duration bursts found in the RHESSI and Swift databases
seem to be represented by different phenomena.

\end{itemize}

\begin{deluxetable}{lccrr}
\tabletypesize{\scriptsize}
\tablecaption{The spectral lags and peak-count rates of the RHESSI GRBs.\label{tab:database}}
\tablewidth{0pt}
\tablehead{
\colhead{GRB\tablenotemark{a}} &
\colhead{Group\tablenotemark{b}} &
\colhead{$L$ (ms)\tablenotemark{c}} &
\colhead{$F$ (s$^{-1}$)\tablenotemark{d}} &
\colhead{$\sigma_F$ (s$^{-1}$)\tablenotemark{e}}
}
\startdata
020214	&	3	&	42.4	$^{+	56.7	}_{-	35.0	}$ &	8885.9	&	221.3	\\
020218	&	3	&	607.0	$^{+	181.9	}_{-	205.4	}$ &	3630.7	&	92.9	\\
020302	&	3	&		                                   &	632.6	&	62.7	\\
020306	&	1	&	1.2	$^{+	15.7	}_{-	17.3	}$ &	9003.0	&	867.4	\\
020311	&	3	&	641.9	$^{+	570.7	}_{-	519.2	}$ &	1571.9	&	119.7	\\
020313	&	3	&		                                   &	891.2	&	89.3	\\
020315	&	3	&		                                   &	504.1	&	91.6	\\
020331	&	3	&		                                   &	307.2	&	71.8	\\
020407	&	3	&		                                   &	775.9	&	72.4	\\
020409	&	3	&		                                   &	268.6	&	56.0	\\
020413	&	3	&		                                   &	1153.5	&	130.8	\\
020417	&	3	&		                                   &	846.8	&	77.1	\\
020418	&	3	&	108.6	$^{+	94.2	}_{-	93.1	}$ &	5618.8	&	289.6	\\
020426	&	1	&		                                   &	3873.5	&	436.2	\\
020430	&	3	&		                                   &	1209.8	&	100.9	\\
020509	&	3	&		                                   &	2785.5	&	310.0	\\
020524	&	3	&		                                   &	958.1	&	116.7	\\
020525A	&	3	&	452.4	$^{+	502.2	}_{-	2154.8	}$ &	975.4	&	159.5	\\
020525B	&	1	&		                                   &	3265.2	&	526.6	\\
020527	&	2	&		                                   &	1467.1	&	278.8	\\
020602	&	3	&		                                   &	2643.6	&	503.4	\\
020603	&	3	&		                                   &	6473.9	&	1042.8	\\
020604	&	3	&		                                   &	1187.1	&	98.9	\\
020620	&	3	&		                                   &	1816.7	&	260.3	\\
020623	&	3	&		                                   &	962.8	&	169.9	\\
020630	&	3	&		                                   &	997.8	&	94.1	\\
020702	&	3	&		                                   &	803.6	&	98.2	\\
020708	&	3	&		                                   &	383.9	&	45.8	\\
020712	&	3	&		                                   &	724.5	&	115.8	\\
020715A	&	1	&		                                   &	2449.4	&	397.6	\\
020715B	&	3	&	135.7	$^{+	55.8	}_{-	47.9	}$ &	10359.6	&	247.5	\\
020725	&	3	&		                                   &	3985.2	&	469.3	\\
020801	&	3	&		                                   &	1740.8	&	384.4	\\
020819A	&	2	&	170.6	$^{+	127.4	}_{-	109.2	}$ &	2235.2	&	353.0	\\
020819B	&	3	&		                                   &	1014.4	&	96.3	\\
020828	&	1	&	6.3	$^{+	84.3	}_{-	46.7	}$ &	5356.8	&	862.4	\\
020910	&	3	&		                                   &	1792.6	&	215.3	\\
020914	&	3	&		                                   &	1284.7	&	268.7	\\
020926	&	3	&		                                   &	618.9	&	63.2	\\
021008A	&	3	&	12.5	$^{+	17.6	}_{-	16.2	}$ &	54724.6	&	1415.5	\\
021008B	&	3	&		                                   &	556.5	&	108.6	\\
021011	&	3	&		                                   &	771.1	&	157.9	\\
021016	&	3	&		                                   &	647.3	&	98.0	\\
021020	&	3	&		                                   &	2275.6	&	125.7	\\
021023	&	3	&		                                   &	1787.9	&	270.7	\\
021025	&	3	&		                                   &	645.3	&	91.9	\\
021102	&	3	&		                                   &	1796.7	&	112.4	\\
021105	&	3	&		                                   &	818.0	&	163.6	\\
021108	&	3	&		                                   &	1072.2	&	87.1	\\
021109	&	3	&		                                   &	903.3	&	85.7	\\
021113	&	3	&		                                   &	374.3	&	89.6	\\
021115	&	3	&		                                   &	484.9	&	89.2	\\
021119	&	3	&	-393.8	$^{+	3000.4	}_{-	2414.5	}$ &	2129.7	&	277.4	\\
021125	&	3	&		                                   &	1434.5	&	174.4	\\
021201	&	1	&	10.5	$^{+	17.9	}_{-	22.5	}$ &	9868.9	&	1044.7	\\
021205	&	3	&		                                   &	579.2	&	39.4	\\
021206	&	3	&	8.9	$^{+	4.0	}_{-	4.2	}$ &	78241.5	&	3719.6	\\
021211	&	3	&		                                   &	1649.6	&	409.6	\\
021214	&	3	&		                                   &	279.9	&	44.6	\\
021223	&	3	&		                                   &	1169.7	&	338.0	\\
021226	&	1	&	16.3	$^{+	21.1	}_{-	27.9	}$ &	5997.7	&	403.9	\\
030102	&	3	&		                                   &	1125.4	&	99.1	\\
030103	&	3	&		                                   &	245.3	&	61.2	\\
030105	&	2	&	6.7	$^{+	28.4	}_{-	22.7	}$ &	4873.4	&	465.0	\\
030110	&	1	&		                                   &	2791.2	&	714.4	\\
030115A	&	3	&	328.0	$^{+	2334.8	}_{-	2009.9	}$ &	1433.8	&	145.1	\\
030115B	&	3	&		                                   &	970.4	&	214.2	\\
030127	&	3	&		                                   &	779.9	&	133.1	\\
030204	&	3	&	196.0	$^{+	303.8	}_{-	349.4	}$ &	2093.2	&	68.3	\\
030206	&	1	&	-6.6	$^{+	56.8	}_{-	16.8	}$ &	3982.8	&	553.3	\\
030212	&	3	&		                                   &	726.2	&	175.5	\\
030214	&	3	&		                                   &	2164.6	&	168.9	\\
030216	&	3	&		                                   &	530.3	&	93.7	\\
030217	&	3	&		                                   &	5531.9	&	548.3	\\
030222	&	3	&		                                   &	772.4	&	89.4	\\
030223	&	3	&	1108.2	$^{+	1740.4	}_{-	1544.0	}$ &	857.3	&	87.0	\\
030225	&	3	&	1338.1	$^{+	4066.8	}_{-	3979.4	}$ &	566.8	&	67.7	\\
030227	&	3	&	-1374.2	$^{+	5702.5	}_{-	4829.1	}$ &	358.0	&	69.8	\\
030228	&	3	&		                                   &	1281.8	&	147.4	\\
030301	&	3	&		                                   &	381.6	&	93.5	\\
030306	&	3	&		                                   &	2740.5	&	135.1	\\
030307	&	3	&	248.3	$^{+	63.8	}_{-	61.4	}$ &	7542.7	&	302.7	\\
030320A	&	3	&	358.1	$^{+	1341.7	}_{-	1042.7	}$ &	1782.3	&	198.9	\\
030320B	&	3	&	415.8	$^{+	1825.9	}_{-	1203.6	}$ &	622.6	&	29.4	\\
030326	&	3	&		                                   &	2034.3	&	116.4	\\
030328	&	3	&		                                   &	891.1	&	141.0	\\
030329A	&	3	&	37.5	$^{+	89.9	}_{-	97.7	}$ &	11876.3	&	219.7	\\
030329B	&	3	&		                                   &	531.8	&	49.5	\\
030331	&	3	&		                                   &	1148.8	&	221.9	\\
030406	&	3	&	240.6	$^{+	168.9	}_{-	169.7	}$ &	6067.2	&	283.3	\\
030410	&	2	&	23.1	$^{+	64.6	}_{-	88.5	}$ &	2392.8	&	308.2	\\
030413	&	3	&	-257.9	$^{+	5958.8	}_{-	1027.3	}$ &	1181.4	&	109.1	\\
030414	&	3	&	1171.0	$^{+	316.2	}_{-	258.2	}$ &	2599.3	&	96.8	\\
030419	&	3	&		                                   &	7826.4	&	226.3	\\
030421	&	3	&	69.0	$^{+	166.4	}_{-	199.6	}$ &	4489.0	&	525.3	\\
030422	&	3	&		                                   &	700.1	&	102.2	\\
030428	&	3	&	24.9	$^{+	22.0	}_{-	27.6	}$ &	5294.8	&	190.8	\\
030501A	&	3	&		                                   &	1961.3	&	146.9	\\
030501B	&	3	&		                                   &	404.4	&	96.2	\\
030501C	&	2	&		                                   &	3029.0	&	600.2	\\
030505A	&	3	&		                                   &	584.9	&	112.9	\\
030505B	&	3	&	-267.8	$^{+	749.1	}_{-	729.4	}$ &	1448.6	&	46.9	\\
030506	&	3	&		                                   &	1315.8	&	186.0	\\
030518A	&	3	&	81.2	$^{+	81.8	}_{-	107.4	}$ &	7982.3	&	640.6	\\
030518B	&	3	&		                                   &	1867.0	&	248.6	\\
030519A	&	3	&		                                   &	1890.9	&	318.7	\\
030519B	&	3	&	17.0	$^{+	17.2	}_{-	19.5	}$ &	15592.7	&	299.0	\\
030523	&	1	&		                                   &	2828.1	&	485.7	\\
030528	&	3	&		                                   &	423.7	&	69.7	\\
030601	&	3	&	465.6	$^{+	679.7	}_{-	929.7	}$ &	1051.9	&	98.4	\\
030614	&	3	&		                                   &	898.5	&	32.7	\\
030626	&	3	&	333.6	$^{+	714.2	}_{-	772.6	}$ &	1434.1	&	78.0	\\
030703	&	3	&		                                   &	229.4	&	54.0	\\
030706	&	3	&		                                   &	1021.3	&	101.4	\\
030710	&	3	&		                                   &	1114.7	&	97.7	\\
030714	&	3	&		                                   &	1499.3	&	129.3	\\
030716	&	3	&		                                   &	377.2	&	68.3	\\
030721	&	3	&	59.9	$^{+	659.6	}_{-	318.1	}$ &	12474.1	&	679.8	\\
030725	&	3	&		                                   &	1014.1	&	93.0	\\
030726A	&	3	&	57.8	$^{+	166.1	}_{-	129.2	}$ &	2005.1	&	94.7	\\
030726B	&	3	&		                                   &	400.9	&	51.9	\\
030728	&	3	&		                                   &	655.4	&	127.2	\\
030824	&	3	&		                                   &	509.0	&	86.0	\\
030827	&	3	&	18.4	$^{+	74.2	}_{-	54.8	}$ &	3678.2	&	355.2	\\
030830	&	3	&		                                   &	1835.5	&	87.4	\\
030831	&	3	&		                                   &	1903.3	&	123.3	\\
030919	&	3	&		                                   &	828.7	&	143.0	\\
030921	&	3	&		                                   &	2414.9	&	127.7	\\
030922A	&	3	&		                                   &	1418.7	&	111.6	\\
030922B	&	3	&	38.7	$^{+	164.4	}_{-	146.0	}$ &	4042.1	&	171.3	\\
030926	&	1	&		                                   &	2094.0	&	422.3	\\
031005	&	3	&		                                   &	661.9	&	132.6	\\
031019	&	3	&		                                   &	1260.8	&	160.4	\\
031024	&	3	&	218.6	$^{+	394.5	}_{-	421.0	}$ &	3055.6	&	552.4	\\
031027	&	3	&	128.2	$^{+	111.0	}_{-	125.4	}$ &	4102.5	&	113.9	\\
031107	&	3	&	-121.0	$^{+	1243.7	}_{-	1523.6	}$ &	1183.0	&	86.4	\\
031108	&	3	&	88.9	$^{+	69.3	}_{-	82.9	}$ &	5923.9	&	157.4	\\
031111	&	3	&	57.1	$^{+	28.6	}_{-	31.6	}$ &	13882.4	&	445.7	\\
031118	&	1	&		                                   &	4669.8	&	592.7	\\
031120	&	3	&	1147.6	$^{+	637.5	}_{-	1068.6	}$ &	991.2	&	57.3	\\
031127	&	3	&		                                   &	634.4	&	94.5	\\
031130	&	3	&		                                   &	1354.9	&	155.0	\\
031214	&	3	&		                                   &	609.8	&	159.3	\\
031218	&	1	&	53.4	$^{+	137.0	}_{-	48.9	}$ &	4732.5	&	1067.7	\\
031219	&	3	&	334.0	$^{+	589.0	}_{-	350.8	}$ &	3262.1	&	439.7	\\
031226	&	3	&		                                   &	640.4	&	72.5	\\
031226	&	3	&		                                   &	612.8	&	86.8	\\
040102	&	3	&		                                   &	1160.1	&	139.1	\\
040108	&	3	&		                                   &	395.7	&	70.3	\\
040113	&	3	&		                                   &	1183.6	&	193.7	\\
040115	&	3	&		                                   &	473.1	&	84.0	\\
040125	&	3	&		                                   &	528.6	&	111.7	\\
040205A	&	3	&		                                   &	497.4	&	152.7	\\
040205B	&	3	&		                                   &	378.0	&	76.3	\\
040207	&	3	&	23.6	$^{+	91.0	}_{-	84.4	}$ &	3375.9	&	115.9	\\
040211	&	3	&		                                   &	751.8	&	181.5	\\
040215	&	3	&		                                   &	391.7	&	46.9	\\
040220	&	3	&	1041.1	$^{+	1167.5	}_{-	1171.1	}$ &	1714.0	&	178.9	\\
040225A	&	3	&		                                   &	576.5	&	104.0	\\
040225B	&	3	&		                                   &	642.5	&	90.0	\\
040228	&	3	&	19.2	$^{+	39.5	}_{-	36.0	}$ &	11483.8	&	261.6	\\
040302A	&	3	&		                                   &	712.0	&	98.2	\\
040302B	&	3	&	101.8	$^{+	47.2	}_{-	32.4	}$ &	13904.2	&	283.4	\\
040303	&	3	&		                                   &	403.8	&	86.2	\\
040312	&	1	&	2.7	$^{+	27.4	}_{-	14.8	}$ &	7197.9	&	690.1	\\
040316	&	3	&	-161.3	$^{+	316.9	}_{-	380.0	}$ &	4356.0	&	452.4	\\
040323	&	3	&		                                   &	650.8	&	174.9	\\
040324	&	1	&	2.7	$^{+	4.5	}_{-	6.5	}$ &	16984.5	&	1112.0	\\
040327	&	3	&		                                   &	405.8	&	56.2	\\
040329	&	2	&	3.6	$^{+	9.2	}_{-	9.6	}$ &	19974.8	&	849.1	\\
040330	&	3	&		                                   &	584.5	&	98.3	\\
040404	&	3	&		                                   &	1096.6	&	183.6	\\
040413	&	1	&		                                   &	5163.3	&	594.7	\\
040414	&	3	&		                                   &	992.3	&	59.5	\\
040421	&	3	&	98.6	$^{+	68.9	}_{-	75.1	}$ &	7197.3	&	220.2	\\
040423	&	3	&		                                   &	663.4	&	111.5	\\
040425	&	3	&	155.3	$^{+	139.1	}_{-	126.2	}$ &	7424.3	&	843.8	\\
040427	&	3	&		                                   &	674.8	&	112.1	\\
040429	&	3	&		                                   &	582.8	&	85.4	\\
040502A	&	3	&		                                   &	3551.8	&	310.5	\\
040502B	&	3	&		                                   &	602.9	&	48.9	\\
040506	&	3	&		                                   &	807.0	&	61.2	\\
040508	&	3	&		                                   &	242.0	&	64.4	\\
040510	&	3	&		                                   &	1097.3	&	123.3	\\
040513	&	3	&		                                   &	218.9	&	49.2	\\
040526	&	3	&		                                   &	499.1	&	96.2	\\
040528	&	3	&	731.5	$^{+	1067.9	}_{-	1191.3	}$ &	1827.0	&	119.2	\\
040531	&	3	&		                                   &	1024.5	&	75.3	\\
040601	&	3	&		                                   &	401.3	&	93.9	\\
040603A	&	3	&		                                   &	726.3	&	138.6	\\
040603B	&	3	&		                                   &	161.9	&	26.5	\\
040605A	&	3	&		                                   &	800.3	&	224.4	\\
040605B	&	1	&	26.0	$^{+	45.0	}_{-	50.8	}$ &	8657.1	&	1243.4	\\
040605C	&	3	&		                                   &	1271.6	&	87.0	\\
040611	&	3	&	-2267.0	$^{+	2983.7	}_{-	1345.5	}$ &	740.1	&	87.8	\\
040619	&	3	&		                                   &	3308.1	&	480.5	\\
040701	&	3	&	135.0	$^{+	245.2	}_{-	318.9	}$ &	1128.0	&	103.9	\\
040719	&	3	&		                                   &	1445.1	&	177.0	\\
040723	&	3	&		                                   &	4168.6	&	697.7	\\
040731	&	3	&	110.2	$^{+	244.6	}_{-	257.4	}$ &	1803.4	&	91.5	\\
040803	&	3	&		                                   &	267.3	&	39.1	\\
040810	&	3	&		                                   &	1654.2	&	109.7	\\
040818	&	3	&	33.2	$^{+	97.4	}_{-	153.4	}$ &	2681.5	&	225.6	\\
040822	&	2	&		                                   &	2226.7	&	374.5	\\
040824	&	3	&		                                   &	282.2	&	45.8	\\
040921	&	1	&		                                   &	2619.0	&	490.6	\\
040925	&	3	&	-136.0	$^{+	2908.9	}_{-	1745.9	}$ &	1263.8	&	167.1	\\
040926	&	3	&	4.1	$^{+	92.7	}_{-	83.7	}$ &	5855.8	&	207.3	\\
041003	&	3	&		                                   &	735.4	&	182.2	\\
041006	&	3	&		                                   &	1307.7	&	254.1	\\
041007	&	2	&	101.0	$^{+	119.5	}_{-	123.3	}$ &	3301.3	&	238.6	\\
041009	&	3	&		                                   &	1382.6	&	111.5	\\
041010	&	1	&	0.7	$^{+	6.3	}_{-	8.0	}$ &	1838.8	&	616.2	\\
041012	&	3	&		                                   &	276.6	&	44.5	\\
041013A	&	3	&		                                   &	530.3	&	39.7	\\
041013B	&	1	&		                                   &	3412.2	&	421.8	\\
041015	&	3	&		                                   &	1630.8	&	189.7	\\
041016	&	3	&		                                   &	509.0	&	78.5	\\
041018	&	3	&	6774.4	$^{+	6756.0	}_{-	7424.5	}$ &	717.6	&	120.4	\\
041101	&	3	&		                                   &	1036.3	&	167.5	\\
041102	&	3	&		                                   &	1437.3	&	188.3	\\
041107	&	3	&		                                   &	883.0	&	83.3	\\
041116	&	3	&		                                   &	269.7	&	52.1	\\
041117	&	3	&		                                   &	1165.2	&	106.5	\\
041120	&	3	&		                                   &	1167.1	&	123.6	\\
041125	&	3	&	-7.8	$^{+	41.0	}_{-	55.4	}$ &	9489.2	&	168.7	\\
041202	&	3	&	76.3	$^{+	102.6	}_{-	93.7	}$ &	5824.1	&	198.0	\\
041211A	&	3	&	56.3	$^{+	657.3	}_{-	825.0	}$ &	1066.9	&	83.9	\\
041211B	&	3	&		                                   &	5126.6	&	629.1	\\
041211C	&	3	&	-6.0	$^{+	12.9	}_{-	12.9	}$ &	11832.5	&	315.0	\\
041213	&	1	&		                                   &	6872.1	&	669.2	\\
041218	&	3	&		                                   &	364.9	&	54.7	\\
041219	&	3	&		                                   &	560.3	&	76.9	\\
041223	&	3	&	-1.6	$^{+	208.1	}_{-	211.7	}$ &	1567.8	&	91.6	\\
041224	&	3	&		                                   &	285.4	&	73.2	\\
041231	&	2	&	21.6	$^{+	187.2	}_{-	177.7	}$ &	2211.1	&	206.3	\\
050124	&	3	&		                                   &	815.2	&	162.3	\\
050126	&	3	&		                                   &	1477.0	&	79.6	\\
050203	&	3	&		                                   &	2870.6	&	234.3	\\
050213	&	3	&	44.2	$^{+	197.8	}_{-	194.9	}$ &	2174.0	&	101.5	\\
050214	&	3	&		                                   &	390.4	&	60.3	\\
050216	&	1	&	6.0	$^{+	99.4	}_{-	76.5	}$ &	4614.1	&	686.4	\\
050219	&	3	&	410.8	$^{+	321.9	}_{-	186.9	}$ &	4760.8	&	259.7	\\
050311	&	3	&		                                   &	465.1	&	88.4	\\
050312	&	1	&	4.4	$^{+	44.5	}_{-	31.0	}$ &	4439.8	&	419.4	\\
050314	&	3	&		                                   &	2101.0	&	151.0	\\
050320	&	3	&		                                   &	838.7	&	114.5	\\
050321	&	3	&		                                   &	880.5	&	128.4	\\
050326	&	3	&		                                   &	3501.9	&	403.9	\\
050328	&	1	&	-43.5	$^{+	79.4	}_{-	84.1	}$ &	12706.3	&	1069.6	\\
050404	&	3	&	29.1	$^{+	35.6	}_{-	39.0	}$ &	6336.6	&	215.9	\\
050409	&	2	&	-1.7	$^{+	16.9	}_{-	14.1	}$ &	18969.8	&	1936.7	\\
050411	&	3	&		                                   &	813.9	&	125.9	\\
050412	&	3	&		                                   &	1455.7	&	140.8	\\
050429	&	3	&		                                   &	2018.6	&	145.8	\\
050430	&	3	&		                                   &	986.8	&	139.4	\\
050501	&	3	&		                                   &	1697.9	&	308.8	\\
050502	&	2	&		                                   &	1132.2	&	235.6	\\
050509	&	3	&	-63.6	$^{+	397.4	}_{-	494.3	}$ &	2282.4	&	145.7	\\
050516	&	3	&		                                   &	259.7	&	71.5	\\
050525A	&	3	&		                                   &	4853.5	&	259.6	\\
050525B	&	3	&	8.5	$^{+	50.0	}_{-	50.3	}$ &	7311.6	&	214.3	\\
050528	&	3	&		                                   &	421.3	&	48.7	\\
050530	&	3	&		                                   &	1963.1	&	273.3	\\
050531	&	3	&	56.6	$^{+	167.3	}_{-	163.6	}$ &	6082.8	&	197.2	\\
050614	&	3	&		                                   &	377.5	&	58.6	\\
050701	&	3	&		                                   &	1173.5	&	166.1	\\
050702	&	2	&		                                   &	951.9	&	187.8	\\
050703	&	3	&		                                   &	2734.2	&	305.4	\\
050706	&	3	&		                                   &	1620.7	&	168.6	\\
050713A	&	3	&	-233.3	$^{+	1804.9	}_{-	2189.4	}$ &	1450.6	&	244.4	\\
050713B	&	3	&		                                   &	431.1	&	69.1	\\
050715	&	3	&		                                   &	2053.4	&	198.9	\\
050717	&	3	&	110.3	$^{+	151.6	}_{-	220.0	}$ &	2456.3	&	143.1	\\
050726	&	3	&	810.5	$^{+	1751.7	}_{-	1334.3	}$ &	1480.4	&	90.7	\\
050729	&	3	&		                                   &	873.7	&	177.5	\\
050802	&	3	&		                                   &	498.5	&	97.6	\\
050805	&	2	&	-6.2	$^{+	30.0	}_{-	20.0	}$ &	3355.3	&	334.6	\\
050809	&	3	&		                                   &	4598.3	&	699.3	\\
050813	&	3	&		                                   &	810.7	&	110.4	\\
050814	&	1	&		                                   &	5499.5	&	1658.3	\\
050817	&	3	&		                                   &	895.5	&	118.2	\\
050820	&	3	&		                                   &	681.3	&	101.9	\\
050824	&	1	&	1.2	$^{+	3.2	}_{-	4.0	}$ &	7474.8	&	930.5	\\
050825	&	2	&	-82.6	$^{+	132.9	}_{-	71.3	}$ &	4307.4	&	585.3	\\
050902	&	3	&		                                   &	631.9	&	149.5	\\
050923	&	3	&		                                   &	1872.2	&	395.3	\\
051009	&	3	&		                                   &	475.8	&	66.9	\\
051012	&	3	&	462.5	$^{+	443.7	}_{-	207.9	}$ &	3780.0	&	331.8	\\
051021	&	3	&	-97.8	$^{+	716.9	}_{-	495.2	}$ &	2117.7	&	221.7	\\
051031	&	3	&		                                   &	678.7	&	45.4	\\
051101	&	3	&		                                   &	921.2	&	258.0	\\
051103	&	1	&	0.6	$^{+	2.4	}_{-	2.8	}$ &	135199.6	&	10873.8	\\
051109	&	3	&		                                   &	882.0	&	92.0	\\
051111	&	3	&		                                   &	516.1	&	102.1	\\
051117	&	3	&	-224.8	$^{+	2133.4	}_{-	3318.9	}$ &	942.4	&	133.9	\\
051119	&	3	&		                                   &	743.4	&	133.7	\\
051124A	&	3	&		                                   &	1234.2	&	194.2	\\
051124B	&	3	&		                                   &	2715.0	&	152.8	\\
051201A	&	3	&		                                   &	211.7	&	37.0	\\
051201B	&	3	&		                                   &	1000.9	&	170.4	\\
051207	&	3	&		                                   &	2440.2	&	121.2	\\
051211	&	3	&	592.3	$^{+	836.1	}_{-	914.9	}$ &	1573.5	&	69.6	\\
051217	&	3	&		                                   &	458.7	&	81.7	\\
051220A	&	3	&	39.0	$^{+	24.1	}_{-	26.5	}$ &	19918.0	&	658.7	\\
051220B	&	3	&		                                   &	321.2	&	73.1	\\
051221	&	1	&	0.0	$^{+	6.6	}_{-	8.7	}$ &	15280.2	&	1298.8	\\
051222	&	3	&		                                   &	334.4	&	78.5	\\
060101	&	3	&	599.3	$^{+	655.1	}_{-	778.6	}$ &	1647.1	&	94.2	\\
060110	&	3	&		                                   &	445.6	&	107.8	\\
060111	&	3	&	-2346.6	$^{+	2031.1	}_{-	1648.5	}$ &	1079.4	&	46.7	\\
060117	&	3	&		                                   &	1300.6	&	113.7	\\
060121A	&	3	&		                                   &	2980.4	&	277.6	\\
060121B	&	3	&		                                   &	343.0	&	46.7	\\
060123	&	3	&	18.4	$^{+	624.7	}_{-	529.7	}$ &	3965.2	&	184.2	\\
060124	&	3	&		                                   &	713.2	&	76.3	\\
060130	&	3	&		                                   &	2285.1	&	376.4	\\
060203	&	1	&	-23.5	$^{+	46.5	}_{-	16.1	}$ &	6089.8	&	831.6	\\
060217	&	3	&		                                   &	2432.4	&	416.0	\\
060224	&	3	&	-944.0	$^{+	3150.4	}_{-	1486.0	}$ &	851.0	&	109.2	\\
060228	&	3	&	-702.0	$^{+	1896.9	}_{-	1770.9	}$ &	717.4	&	89.8	\\
060303	&	1	&	21.2	$^{+	46.3	}_{-	53.5	}$ &	9343.4	&	999.5	\\
060306	&	3	&	50.3	$^{+	12.5	}_{-	11.0	}$ &	105153.0	&	3272.9	\\
060309	&	3	&		                                   &	355.6	&	62.1	\\
060312A	&	1	&		                                   &	1526.6	&	337.6	\\
060312B	&	3	&		                                   &	299.2	&	71.4	\\
060313	&	3	&		                                   &	602.2	&	103.3	\\
060323	&	3	&	-186.2	$^{+	270.0	}_{-	253.4	}$ &	3754.8	&	169.1	\\
060325	&	3	&	189.3	$^{+	235.1	}_{-	266.9	}$ &	6080.4	&	638.4	\\
060401	&	3	&	110.2	$^{+	268.0	}_{-	207.9	}$ &	2527.5	&	208.8	\\
060408	&	3	&		                                   &	449.7	&	85.1	\\
060415	&	3	&		                                   &	315.7	&	78.6	\\
060418	&	3	&		                                   &	694.3	&	74.4	\\
060421A	&	3	&		                                   &	789.4	&	78.5	\\
060421B	&	3	&	312.3	$^{+	208.1	}_{-	170.2	}$ &	2812.8	&	106.2	\\
060425	&	1	&	5.3	$^{+	4.3	}_{-	4.8	}$ &	2367.7	&	521.1	\\
060428	&	3	&		                                   &	274.9	&	60.2	\\
060429	&	1	&	3.2	$^{+	14.7	}_{-	14.5	}$ &	20278.7	&	2240.6	\\
060505	&	3	&	277.1	$^{+	5726.4	}_{-	1422.9	}$ &	1384.9	&	256.7	\\
060528	&	3	&	2441.1	$^{+	4000.4	}_{-	4289.8	}$ &	656.8	&	77.5	\\
060530	&	3	&		                                   &	889.0	&	143.9	\\
060610	&	1	&	10.0	$^{+	14.0	}_{-	21.1	}$ &	9569.6	&	748.7	\\
060614	&	3	&		                                   &	498.7	&	40.7	\\
060622	&	3	&	-1510.2	$^{+	1932.3	}_{-	1524.6	}$ &	1038.4	&	71.5	\\
060624	&	3	&		                                   &	9648.0	&	610.3	\\
060625	&	3	&		                                   &	1179.5	&	122.8	\\
060630	&	3	&		                                   &	1377.2	&	83.7	\\
060708	&	1	&	-5.6	$^{+	21.8	}_{-	20.4	}$ &	16293.3	&	1641.8	\\
060729	&	3	&		                                   &	588.4	&	100.8	\\
060805	&	3	&	18.3	$^{+	54.8	}_{-	69.6	}$ &	10612.2	&	474.3	\\
060811	&	3	&		                                   &	1742.2	&	103.9	\\
060819	&	3	&		                                   &	949.2	&	159.3	\\
060823	&	2	&		                                   &	811.4	&	167.0	\\
060919	&	3	&		                                   &	368.6	&	61.5	\\
060920	&	3	&	22.4	$^{+	61.9	}_{-	80.2	}$ &	4809.8	&	181.2	\\
060925	&	3	&		                                   &	1840.6	&	114.1	\\
060928	&	3	&	144.7	$^{+	154.7	}_{-	151.5	}$ &	5934.5	&	281.3	\\
061005	&	3	&		                                   &	2187.7	&	239.8	\\
061006A	&	1	&	9.4	$^{+	167.7	}_{-	123.0	}$ &	6404.0	&	940.7	\\
061006B	&	2	&		                                   &	2359.0	&	286.1	\\
061007	&	3	&	66.2	$^{+	170.0	}_{-	201.1	}$ &	3465.1	&	305.1	\\
061012	&	3	&		                                   &	1474.0	&	171.6	\\
061013	&	3	&		                                   &	776.2	&	102.5	\\
061014	&	1	&		                                   &	2308.8	&	291.6	\\
061022	&	3	&		                                   &	392.3	&	77.3	\\
061031	&	3	&		                                   &	1112.8	&	238.5	\\
061101	&	3	&		                                   &	631.7	&	88.3	\\
061108	&	3	&		                                   &	2267.8	&	219.2	\\
061113	&	3	&	36.4	$^{+	67.3	}_{-	89.2	}$ &	3665.0	&	233.6	\\
061117	&	3	&		                                   &	251.6	&	56.0	\\
061121	&	3	&	-57.8	$^{+	216.2	}_{-	184.8	}$ &	3131.4	&	176.9	\\
061123	&	3	&		                                   &	1360.8	&	172.8	\\
061126	&	3	&	194.3	$^{+	166.7	}_{-	194.4	}$ &	2897.4	&	120.5	\\
061128	&	1	&	-3.0	$^{+	14.5	}_{-	19.2	}$ &	15183.0	&	1310.9	\\
061205	&	3	&		                                   &	427.5	&	83.8	\\
061212	&	3	&	13.0	$^{+	19.8	}_{-	21.4	}$ &	9820.5	&	440.1	\\
061222	&	3	&		                                   &	1367.9	&	263.2	\\
061229	&	3	&		                                   &	680.4	&	68.7	\\
061230	&	3	&		                                   &	641.3	&	133.8	\\
070113	&	1	&		                                   &	1986.8	&	377.5	\\
070116	&	3	&		                                   &	736.2	&	111.8	\\
070120	&	3	&		                                   &	395.8	&	70.0	\\
070121	&	3	&		                                   &	178.8	&	49.4	\\
070125	&	3	&	-9.6	$^{+	167.9	}_{-	207.8	}$ &	4841.5	&	149.0	\\
070214	&	3	&		                                   &	425.7	&	98.3	\\
070220	&	3	&		                                   &	972.7	&	103.0	\\
070221	&	3	&		                                   &	511.0	&	103.0	\\
070307	&	3	&		                                   &	502.2	&	56.2	\\
070402	&	3	&		                                   &	829.0	&	113.1	\\
070420	&	3	&		                                   &	509.7	&	87.5	\\
070508	&	3	&		                                   &	1627.4	&	168.9	\\
070516	&	1	&	28.9	$^{+	59.5	}_{-	70.1	}$ &	2376.6	&	268.6	\\
070531	&	3	&		                                   &	149.7	&	34.0	\\
070614	&	1	&	2.4	$^{+	29.2	}_{-	24.8	}$ &	2874.7	&	275.5	\\
070622	&	3	&	39.9	$^{+	57.8	}_{-	42.9	}$ &	3397.0	&	207.5	\\
070626	&	3	&		                                   &	2058.8	&	111.6	\\
070710	&	3	&		                                   &	379.1	&	66.7	\\
070717	&	3	&		                                   &	347.5	&	59.0	\\
070722	&	3	&		                                   &	321.7	&	77.4	\\
070724	&	3	&		                                   &	259.1	&	37.5	\\
070802	&	3	&	15.8	$^{+	48.0	}_{-	58.1	}$ &	2426.5	&	234.0	\\
070817	&	3	&		                                   &	541.2	&	55.5	\\
070819	&	3	&		                                   &	915.7	&	75.6	\\
070821	&	3	&	663.0	$^{+	825.1	}_{-	764.1	}$ &	1956.7	&	77.2	\\
070824	&	2	&	-5.9	$^{+	120.0	}_{-	109.8	}$ &	3276.6	&	284.3	\\
070825	&	3	&	875.1	$^{+	1116.6	}_{-	1635.1	}$ &	1335.1	&	94.6	\\
070917	&	3	&		                                   &	347.2	&	65.7	\\
071013	&	3	&		                                   &	963.7	&	211.9	\\
071014	&	3	&		                                   &	1094.5	&	155.0	\\
071030	&	3	&		                                   &	570.0	&	82.8	\\
071104	&	3	&		                                   &	747.0	&	98.2	\\
071204	&	1	&		                                   &	2901.5	&	297.5	\\
071217	&	3	&		                                   &	814.3	&	143.8	\\
080114	&	3	&	285.7	$^{+	233.2	}_{-	252.3	}$ &	2321.1	&	144.3	\\
080202	&	3	&		                                   &	462.1	&	75.7	\\
080204	&	3	&	60.5	$^{+	292.9	}_{-	244.0	}$ &	1368.6	&	180.0	\\
080211	&	3	&	1139.6	$^{+	2898.5	}_{-	1804.8	}$ &	1349.4	&	143.3	\\
080218	&	3	&		                                   &	404.1	&	85.2	\\
080224	&	3	&		                                   &	1795.1	&	186.7	\\
080318	&	3	&		                                   &	385.3	&	90.3	\\
080319	&	3	&	157.9	$^{+	551.1	}_{-	1112.4	}$ &	885.8	&	111.1	\\
080320	&	3	&	34.0	$^{+	381.6	}_{-	569.9	}$ &	1722.6	&	110.4	\\
080328	&	3	&		                                   &	886.7	&	47.1	\\
080330	&	3	&	26.3	$^{+	495.1	}_{-	350.8	}$ &	1685.0	&	98.1	\\
080408	&	2	&	73.3	$^{+	133.6	}_{-	230.4	}$ &	2546.1	&	353.7	\\
080413	&	3	&		                                   &	544.1	&	141.2	\\
080425	&	3	&		                                   &	674.6	&	120.5	\\
\enddata
\tablenotetext{a}{RHESSI GRB number.}
\tablenotetext{b}{The assignment to the GRB group: 1 - short,\\
                  2 - intermediate, 3 - long.}
\tablenotetext{c}{Spectral lags were calculated from the difference
of the count light-curves at the energy intervals $400-1500$\,keV and $25-120$\,keV.
The errors compose of the 95\,\%~CL statistical uncertainty and the light-curve's time resolution.}
\tablenotetext{d}{Peak-count rates derived in the band $25-1500$\,keV.}
\tablenotetext{e}{One sigma statistical uncertainties of the peak-count rates.}
\end{deluxetable}

\acknowledgments
We wish to thank Z. Bagoly, L.G. Bal\'azs, I. Horv\'ath and P. M\'esz\'aros
for useful discussions and comments on the manuscript.
We also thank to R. Aptekar and T.L. Cline for providing the Konus-Wind data.
Thanks are also due for the valuable remarks of anonymous referees.
This study was supported by the OTKA grant K77795,
by the Grant Agency of the Czech Republic grant No.
P209/10/0734, by the Research Program MSM0021620860 of the Ministry
of Education of the Czech Republic, and by Creative Research Initiatives
(RCMST) of MEST/NRF and the World Class University grant no R32-2008-000-101300.

\appendix
\section{Uncertainties of Decimated Data}
\label{app:1}
This part describes the one sigma uncertainties of the bin counts for the decimated
RHESSI data \citep{curt02,smi02}.

\subsection{Full decimation}
Calculation of one sigma uncertainty $\sigma_{\mathrm C_\mathrm{dc}}$ for
the bin counts $C_\mathrm{dc}$ of fully decimated data and then corrected
for this decimation is the following. For corrected bin counts $C_\mathrm{dc}$
it holds:
\begin{equation}
C_\mathrm{dc}=f_\mathrm{d}.C_\mathrm d,
\end{equation}
where $f_\mathrm{d}$ is the decimation factor (weight), usually equal to 4 or 6 for the
RHESSI data, and $C_\mathrm d$ is the number of counts in a bin of the
decimated signal.
If we assume that counts in a bin follow Poisson statistics then:
\begin{equation}
\label{eq:full-dec}
\sigma_{\mathrm C_\mathrm{dc}}=\left|\frac{\partial C_\mathrm{dc}}{\partial
C_\mathrm{d}}\right|\sigma_{\mathrm C_\mathrm d}=f_\mathrm{d}\sqrt{C_\mathrm
d}=\sqrt{f_\mathrm{d}.C_\mathrm{dc}},
\end{equation} where $\sigma_{\mathrm C_\mathrm{d}}$ is the dispersion of the $C_\mathrm d$.

\subsection{Partial decimation}
Now consider the situation in which counts in a bin are only partially
decimated, i.e. they consists of the non-decimated $C_1$ and the decimated
signal $C_\mathrm{2,d}$. This situation may happen when we sum counts over
the energy band $[E_1;E_2]$, $E_1<E_0<E_2$, and only the counts below the
energy $E_0$ are decimated. Then the corrected signal $C_{dc}$ is equal to:
\begin{equation}
C_{dc}=C_1+C_\mathrm{2,dc}=C_1+f_\mathrm{d}.C_\mathrm{2,d},
\end{equation}
where $f_\mathrm{d}$ is again the decimation factor, and $C_\mathrm{2,dc}$ is corrected
part of the signal that was decimated.
One sigma uncertainty $\sigma_\mathrm{C_{dc}}$ is then equal to:
\begin{eqnarray}
\label{eq:part-dec}
\nonumber
\sigma_\mathrm{C_{dc}}=\sqrt{\left(\frac{\partial C}{\partial
C_1}\right)^2\sigma_{C_1}^2+\left(\frac{\partial C}{\partial
C_\mathrm{2,d}}\right)^2\sigma_{C_\mathrm{2,d}}^2}=
\\
=\sqrt{\sigma^{2}_{\mathrm{C}_1}+f_\mathrm{d}^2.\sigma^{2}_{\mathrm
C_\mathrm{2,d}}}=\sqrt{C_1+f_\mathrm{d}^2.C_\mathrm{2,d}}=\sqrt{C_1+f_\mathrm{d}.C_\mathrm{2,dc}},
\end{eqnarray}
where $\sigma_{C_1}=\sqrt{C_1}$ is the dispersion of the non-decimated part
of the bin counts and $\sigma_{C_\mathrm{2,d}}=\sqrt{C_\mathrm{2,d}}$ is the
dispersion of the decimated part of the bin counts.

\end{document}